\documentclass[aps,prd,twocolumn,showpacs,groupedaddress,showkeys,preprintnumbers]{revtex4}
\usepackage{amsmath}
\usepackage{amssymb}
\usepackage{graphicx}
\usepackage{subfigure}
\usepackage{color}
\usepackage[bookmarksnumbered, bookmarksopen, breaklinks, colorlinks]{hyperref}
\newcommand{\be}{\begin{equation}}
\newcommand{\ee}{\end{equation}}

\newcommand{\bea}{\begin{eqnarray}}
\newcommand{\eea}{\end{eqnarray}}
\newcommand{\bean}{\begin{eqnarray*}}
\newcommand{\eean}{\end{eqnarray*}}
\newcommand{\e}{{\rm e}}

\newcommand{\n}[1]{\label{#1}}

\begin{document}

\title{Coherent network analysis technique for discriminating
gravitational-wave bursts from instrumental noise}

\date{\today}

\author{Shourov Chatterji}
\author{Albert Lazzarini}
\author{Leo Stein}
\author{Patrick J. Sutton}
\affiliation{LIGO - California Institute of Technology, Pasadena, CA 91125}

\author{Antony Searle}
\affiliation{Australian National University, Canberra, ACT 0200, Australia}

\author{Massimo Tinto}
\affiliation{Jet Propulsion Laboratory, California Institute of Technology, Pasadena, CA 91109}

\begin{abstract}
  Existing coherent network analysis techniques for detecting
  gravitational-wave bursts simultaneously test data from multiple
  observatories for consistency with the expected properties of the
  signals.  These techniques assume the output of the detector network
  to be the sum of a stationary Gaussian noise process and a
  gravitational-wave signal, and they may fail in the presence of
  transient non-stationarities, which are common in real detectors.
  In order to address this problem we introduce a 
  consistency test that is robust against noise non-stationarities and
  allows one to distinguish between gravitational-wave bursts and
  noise transients.  This technique does not require any {\em a priori}
  knowledge of the putative burst waveform. 
\end{abstract}



\preprint{LIGO-P060009-01-E}

\maketitle

\section{Introduction}
\label{sec:introduction}

Gravitational-wave bursts (GWBs) are among the most exciting classes
of signals that large-scale, broad-band interferometric 
gravitational-wave observatories will attempt to detect. These instruments have
already started to collect data, and are beginning to coordinate their
activities in order to perform joint, world-wide, searches. The United
States LIGO \cite{LIGO} observatory has reached its design sensitivity
goal, and is currently conducting its fifth data-taking run. Both the
British-German GEO600 \cite{GEO} interferometer and the Japanese TAMA
\cite{TAMA} detector have conducted multiple data-taking runs, often
in coincidence with LIGO.  The French-Italian VIRGO \cite{VIRGO}
detector is approaching its operational phase with an anticipated
sensitivity comparable to that of the US detectors. 

\begin{figure}[!htp]
\centering
\resizebox{\columnwidth}{!}{\includegraphics{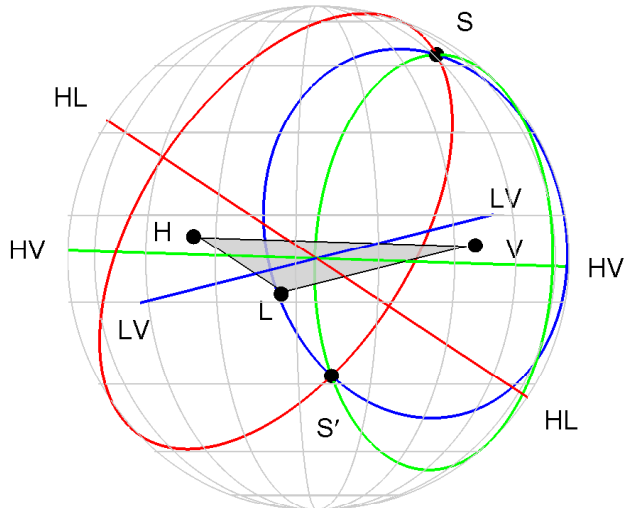}}
\caption{Geometry of the network and travel times spent by a GWB to
  propagate across a three-detector network (detectors ``H'', ``L'', and ``V''). The locus of constant time delay
  between two detectors form a ring on the sky concentric about the baseline
  between the two sites.  For three detectors, these rings may
  intersect in two locations.  One is the true source direction, $S$,
  while the other ($S^{\prime}$) is its mirror image with respect to
  the geometrical plane passing through the three sites.  This
  two-fold ambiguity can be resolved by further considering the
  amplitudes of the responses.  For four or more detectors there is 
  a unique intersection point of all of the rings.}
\label{fig:ringgeometry}
\end{figure}

Potential sources of GWBs include merging compact objects
\cite{FlHu:98a,FlHu:98b,Pr:05}, core-collapse supernovae
\cite{ZwMu:97,DiFoMu:02b,OtBuLiWa:04,ShSe:04}, and gamma-ray burst
engines \cite{Me:02}. By considering the anticipated strengths of the
gravitational-wave bursts emitted by these systems and the present
performance of the detectors, it is clear that coincident experiments
are necessary to maximize the chances of successfully identifying an
astronomical event.  This is because signals emitted by these sources
cannot currently be modeled with sufficient accuracy to distinguish
them from transient noise non-stationarities (``glitches'') affecting
the data of the detectors.  In addition, a real burst signal may not
be much larger than the noise levels in the detectors.  More detectors
at the same site provide one way for increasing the signal-to-noise
ratio (SNR), but a network of detectors spaced across the globe
inherently provides much more information about the wave and its
source. With a network of three or more interferometers, for instance,
one has in principle sufficient information for good SNRs to infer the
direction to the source. If the events are not corroborated by an
electromagnetic detection (optical, X-ray, or radio) this could be
crucial. Such a network also has enough information to reconstruct the
two independent polarization amplitudes of the wave. This is possible
because three interferometers provide three independent measures of
the gravitational wave (which are functions of time) and two
independent time delays. If we consider the geometrical plane passing
through the three sites of the interferometers, the two independent
time delays jointly identify two possible points in the sky where the
signal could have come from that are mirror-images of each other with
respect to this plane; see Figure~\ref{fig:ringgeometry}. Since the
detector antenna patterns are not symmetric with respect to this
plane, it is further possible to resolve this two-fold ambiguity by
properly accounting for the antenna pattern asymmetry in the analysis
of the data.  Once the sky position has been determined it is
straightforward to make a minimum-variance estimate of the two
polarization waveforms of the GWB as linear combinations of the
detector data streams.  The extraction of this information from the
combined responses of the individual detectors of the network is
called the ``solution of the inverse problem'' in gravitational-wave
astronomy.

The inverse problem for gravitational-wave bursts with a network of
three wide-band, widely separated detectors was first solved by
G\"ursel and Tinto \cite{GuTi:89}. Their technique, which is referred
to in recent literature as the {\it null-stream} method, relies on the
observation that a gravitational-wave burst present in the data of a
network of three wide-band detectors must satisfy a unique closure
condition.  G\"ursel and Tinto studied a two-parameter family of
linear combinations of the three data sets, in which the two
parameters correspond to the two angular coordinates of the
hypothesized sky location of the source.  They showed that when the
two parameters coincide with the true location of the source then the
gravitational-wave burst is canceled precisely in the linear
combination.  This point is located by applying a least-squares
minimization (i.e., a $\chi^2$ test) to the linear combination.  In
\cite{GuTi:89} it was also shown that this condition holds regardless
of the time dependence of the two polarization waveforms of the burst.
This remarkable result makes this method very powerful for solving the
inverse problem since it does not require {\em a priori\/} knowledge
of the burst waveforms.

Even with the G\"ursel-Tinto technique, lack of knowledge of the 
signal waveforms presents a serious impediment in searches for
GWBs.  This is because gravitational-wave
detectors exhibit transient noise fluctuations, or ``glitches''.
Without a model for GWB waveforms, it is not obvious how to determine
with confidence whether a candidate detection is a real
gravitational-wave burst signal or a ``false alarm'' due to noise
fluctuations occurring in coincidence in the detectors of the network.
A confident detection of gravitational-wave bursts requires the
ability to distinguish them from such noise transients. In this
context, as pointed out in \cite{TAE00,TL04,AEI05}, the null-stream
method could also be used for discriminating gravitational-wave bursts
from noise-triggered fluctuations affecting the data of the detectors.
Put simply, uncorrelated noise glitches should not cancel in the 
null stream.  In principle, therefore, they can be vetoed by setting 
a threshold on the maximum allowable $\chi^2$ value.  In practice, 
however, a $\chi^2$ veto test is vulnerable to effects that prevent 
precise cancellation of strong GWBs, such as calibration errors, 
and may be ineffective for weak glitches, for which there are 
often sky positions which yield $\chi^2 \sim 1$ per degree of freedom 
even though the glitches are uncorrelated.  A $\chi^2$ threshold which is low 
enough to pass a GWB with poor data calibration may also pass a weak 
glitch.  A $\chi^2$ threshold which is high enough to reject the 
glitch may also reject the GWB.

In this paper we propose a modified null-stream based technique for 
discriminating GWBs from noise glitches.  This technique is based 
on comparing the energy in the null stream to that expected if the 
transients in the detectors are uncorrelated.  This second energy 
measure, which we call the {\em incoherent energy}, provides an 
effective measure of the significance of the $\chi^2$ test and 
renders it robust against both calibration uncertainties and weak 
glitches.  Like the G\"ursel-Tinto analysis for determining 
the source direction, our null-stream consistency test does 
not require any {\em a priori} knowledge of the GWB (or glitch) 
waveforms.

Our paper is organized as follows.  In Section \ref{SECII} we show
that there exists a very general and elegant procedure for deriving
the null stream for an arbitrary number of detectors with colored
noise.  We then address the issue that the null-stream $\chi^2$ alone can not
reliably distinguish between a gravitational-wave burst and 
noise-generated glitches.  This is done by introducing a complementary 
energy measure, the incoherent energy, and demonstrating that 
GWBs and glitches separate in the two-dimensional space of null 
and incoherent energies.  This allows us to
identify (or ``veto'') noise-generated events and hence make the
null-stream analysis robust against glitches.

In Section \ref{SECIII} we discuss the results of the numerical
simulation of our statistical test applied to the LIGO-Virgo
3-detector network.  We assume the three interferometers to be working
at the LIGO design sensitivity, and quantify the ability of our method
to distinguish true GWBs from coincident noise glitches.  Although our
numerical implementation is not optimized for the signals under
consideration, it indicates that gravitational-wave bursts observed in
each detector with SNRs of about $10-20$ can reliably be distinguished
from noise glitches of similar energy, and that a significant improvement
over the statistics based on the null-stream $\chi^2$ alone is
achieved.  Our conclusions and future work plans are presented in
Section \ref{SECIV}.

\section{Analysis}
\label{SECII}

Three or more detectors provide redundant measurements of the two
polarization components $h_+$, $h_\times$ of a gravitational wave.  It
is therefore possible to construct linear combinations of the data
streams that do not contain any gravitational-wave component, i.e.,
that consist only of detector noise.  In this section we derive these
linear combinations, known as ``null streams,'' for networks
containing an arbitrary number of detectors whose noises are different
and colored.

\subsection{Conventions}

The conventions used for the notation in this report are described in
Table~\ref{tab:notation}.
\begin{table*}[tbp]
\begin{center}
\begin{tabular}{|l l|}
\hline
$D$
  &   Number of detectors in the network. \\
$\alpha,\beta \in [1,\ldots,D]$
  &  Index specifying detector.  \\
$\boldsymbol{X}$ 
  &  (Boldface) A vector or matrix on the space of detectors. \\
$\boldsymbol{X}^T$ 
  &  Matrix transpose of $\boldsymbol{X}$. \\
$N$
  &  Number of data samples from each detector being analyzed.  \\
$j,k \in [0,\ldots,N-1]$
  &  Index specifying time or frequency sample.  \\
$h_+,h_\times$
  &  ``Plus'' and ``cross'' polarization waveforms of the gravitational wave.  \\
$\widehat{\Omega}_s$
  &  Sky position of the gravitational-wave source. \\
$\widehat{\Omega}$
  &  Trial sky position. \\
$\boldsymbol{d}, \boldsymbol{n}, \boldsymbol{F}$
  &  Data, noise, and antenna responses in strain. \\
$\boldsymbol{d_{\textrm{w}}}, \boldsymbol{n_{\textrm{w}}}, \boldsymbol{F_{\textrm{w}}}$
  &  Noise-weighted (whitened) data, noise, and antenna responses. \\
\hline
\end{tabular}
\caption{\label{tab:notation}
    Notation conventions for commonly used quantities in this paper.
}
\end{center}
\end{table*}

For a plane \footnote{Any gravitational wave at the Earth, produced by
  a source at astronomical distances, can be regarded as a plane
  wave.} gravitational wave incident from a direction
$\widehat{\Omega}_s$ the strain $h_{+,\times}$ at the position
$\vec{r}_\alpha$ is related to that at some (arbitrary) reference
position $\vec{r}_0$ by 
\be 
h_{+,\times}(t+\Delta t_\alpha(\widehat{\Omega}_s),\vec{r}_\alpha)
  =  h_{+,\times}(t,\vec{r}_0) \, , 
\ee 
\noindent
where the time delay $\Delta t_\alpha(\widehat{\Omega}_s)$ is given by 
\be\n{eqn:delay} 
\Delta t_\alpha(\widehat{\Omega}_s)
  \equiv  \frac{1}{c}(\vec{r}_0-\vec{r}_\alpha)\cdot\widehat{\Omega}_s \, .  
\ee
\noindent
We can therefore compare the gravitational-wave signals measured by
detectors at different locations by shifting the time-series data from
each detector according to (\ref{eqn:delay}), provided we know the sky
location of the source.

The time-series signal produced in detector $\alpha$ at
$\vec{r}_\alpha$ by a gravitational-wave $h_{+,\times}$ incident from
a sky position $\widehat{\Omega}_s$ is
\bea\n{eqn:decomposition}
d_\alpha(t+\Delta t_\alpha(\widehat{\Omega}_s))
 & = &  F^+_\alpha(\widehat{\Omega}_s) h_+(t)
        + F^\times_\alpha(\widehat{\Omega}_s) h_\times(t)
\nonumber \\
&   &  \mbox{} + n_\alpha(t+\Delta t_\alpha(\widehat{\Omega}_s)) \, .
\eea
\noindent
Here $n_\alpha$ is the stationary background noise of
detector $\alpha$ as a function of time, calibrated to units of
strain, and $F^+_\alpha, F^\times_\alpha$ are the antenna response
functions \cite{GuTi:89} for detector $\alpha$ for the sky position
$\widehat{\Omega}_s$ of the gravitational wave source.  For brevity,
we write $h_{+,\times}(t) \equiv h_{+,\times}(t,\vec{r}_0)$.

\subsection{Null stream construction}
\label{sec:analysis:nullstream}

Let us assume for the moment that we know the direction
$\widehat{\Omega}_s$ to the source.  Then we can time-shift the data
from each detector as in \eqref{eqn:decomposition}, and drop explicit
references to the time delays $\Delta t_\alpha$ and the sky position.  Transforming to the
Fourier domain, equation \eqref{eqn:decomposition} becomes
\begin{equation}
\tilde{d}_\alpha(f)
  =  F^+_\alpha \tilde{h}_+(f) + F^\times_\alpha \tilde{h}_\times(f)
     + \tilde{n}_\alpha(f) \ ,
\end{equation}
The Fourier transform and its inverse are defined by
\begin{eqnarray}
\tilde{d}_\alpha(f)
  & = &  \int_{- \infty}^{+ \infty} \!\!\! 
         d_\alpha(t) \ e^{- 2 \pi i f t}
  \, dt \, , \nonumber \\
{d}_\alpha(t)
  & = &  \int_{- \infty}^{+ \infty} \!\!\! 
         \tilde{d}_\alpha(f) \ e^{2 \pi i f t} \, df
         \, \ .
         \label{Fourier}
\end{eqnarray}
The one-sided strain noise power spectral density $S_{\alpha}(f)$ of
the stationary noise $n_\alpha$ is given by
\be\n{eq:strainnoise}
\langle 
    \tilde{n}_\alpha(f)
    \tilde{n}_\beta^{*}(f') 
\rangle
  =  \frac{1}{2} \delta_{\alpha \beta} \delta(f-f') S_\alpha(f) \, .
\ee
\noindent
For present detectors $S_\alpha(f)$ is a strongly varying function of
frequency.  Since it will prove convenient to work with white-noise
data, without loss of generality we divide the strain data at each
frequency by the estimated amplitude spectrum $\sqrt{S_\alpha(f)/2}$ of
the corresponding detector noise. The whitened data $\tilde{d}_{w\alpha}$
is then given by
\bea\n{eq:d}
\tilde{d}_{w\alpha}(f)
  & \equiv & \frac{\tilde{d}_\alpha(f)}{\sqrt{S_\alpha (f)/2}} \nonumber \\
  & = &  F^+_{w\alpha} \tilde{h}_+(f) 
         + F^\times_{w\alpha} \tilde{h}_\times(f) 
         + \tilde{n}_{w\alpha}(f) \, ,
\eea
where the $n_{w\alpha}(t)$ are unit Gaussian noise processes and 
$F^{+,\times}_{w\alpha}$ are the noise-weighted antenna responses
\be
F^{+,\times}_{w\alpha}(\widehat{\Omega}_s,f) \equiv
\frac{F^{+,\times}_\alpha(\widehat{\Omega}_s)}{\sqrt{S_\alpha (f)/2}} \, .
\ee
\noindent
The $F^{+,\times}_{w\alpha}$ contain all of the information on the detector
sensitivity, both as functions of frequency and source sky
position.

For a network of $D$ detectors, equation \eqref{eq:d} can be written
in the equivalent matrix form
\begin{equation}
\left[
\begin{array}{c}
  \tilde{d}_{w1} \\
  \tilde{d}_{w2} \\
  \vdots \\
  \tilde{d}_{wD}
\end{array}
\right] = \left[
\begin{array}{cc}
  F_{w1}^+ & F_{w1}^\times \\
  F_{w2}^+ & F_{w2}^\times \\
  \vdots & \vdots\\
  F_{wD}^+ & F_{wD}^\times
\end{array}
\right]\left[
\begin{array}{c}
  \tilde{h}_+ \\
  \tilde{h}_\times
\end{array}
\right]+\left[
\begin{array}{c}
  \tilde{n}_{w1} \\
  \tilde{n}_{w2} \\
  \vdots \\
  \tilde{n}_{wD}
\end{array}
\right] \ ,
\end{equation}
or 
\be\label{eqn:matrixform}
\boldsymbol{\widetilde{d}}_{\textrm{w}}
  =  \boldsymbol{F_\textrm{w}} \, \boldsymbol{\widetilde{h}}
     + \boldsymbol{\widetilde{n}_\textrm{w}} \, ,
\ee
where we use boldface to denote vectors and matrices.  Here 
\be\n{eqn:hmatrix}
\boldsymbol{\widetilde{h}} \equiv \left[
\begin{array}{c}
  \tilde{h}_+ \\
  \tilde{h}_\times
\end{array} \right] \, ,
\ee
and the matrix $\boldsymbol{F_\textrm{w}}$ is defined as
\be
\boldsymbol{F_\textrm{w}}(\widehat{\Omega}_s, f)
  \equiv
     \left[
         \begin{array}{cc}
             \boldsymbol{F_\textrm{w}^+} & \boldsymbol{F_\textrm{w}^\times}
         \end{array}
     \right]
  =  \left[
         \begin{array}{cc}
             F_{\textrm{w}1}^+ & F_{\textrm{w}1}^\times \\
             F_{\textrm{w}2}^+ & F_{\textrm{w}2}^\times \\
             \vdots & \vdots\\
             F_{\textrm{w}D}^+ & F_{\textrm{w}D}^\times
         \end{array}
     \right] \, .
\ee
This form makes it clear that, regardless of the functional form of
$\tilde{h}_{+,\times}$, the gravitational-wave burst can only
contribute to the network output along the directions
$\boldsymbol{F_\textrm{w}^+}$ and $\boldsymbol{F_\textrm{w}^\times}$.  The
construction of null streams is thus obvious: we simply project the
data $\boldsymbol{{\widetilde{d}}}$ orthogonally to these directions.
Formally, we select a new orthonormal Cartesian coordinate basis
$\boldsymbol{e}_i$ for the space of
$\boldsymbol{{\widetilde{d}}}$ in which vectors
$\boldsymbol{e}_{D-1}$ and $\boldsymbol{e}_D$ span
$\boldsymbol{F_\textrm{w}^+}$ and $\boldsymbol{F_\textrm{w}^\times}$.  The
remaining basis vectors $\boldsymbol{e}_1, \ldots,
\boldsymbol{e}_{D-2}$ are then orthogonal to
$\boldsymbol{F_\textrm{w}^+}$ and $\boldsymbol{F_\textrm{w}^\times}$,
\be
\boldsymbol{F_\textrm{w}^+} \cdot \boldsymbol{e}_i = 0
= \boldsymbol{F_\textrm{w}^\times} \cdot \boldsymbol{e}_i \qquad i\in\{1,\ldots,D-2\} \, .
\ee
We say that the $\boldsymbol{e}_{i=1,\ldots,D-2}$ form an
orthonormal basis for the {\em null space} of $\boldsymbol{F_\textrm{w}}^T$
\footnote{The null space of $\boldsymbol{F_\textrm{w}}^T$ is the set of all
  vectors $\boldsymbol{x}$ such that
  $\boldsymbol{F_\textrm{w}}^T\boldsymbol{x}=0$.}, hence the term 
``null stream formalism.''  We then construct a $(D-2)
\times D$ matrix $\boldsymbol{A}$ whose rows are the components of
this orthonormal basis, 
\begin{equation}
\boldsymbol{A}(\widehat{\Omega}_s, f) \equiv
    \left[
        \begin{array}{c}
          \boldsymbol{e}_1^T \\
          \vdots \\
          \boldsymbol{e}_{D-2}^T
        \end{array}
    \right] \, .
\end{equation}
\noindent
By construction $\boldsymbol{A}$ is orthogonal to 
$\boldsymbol{F_\textrm{w}^+}$ and $\boldsymbol{F_\textrm{w}^\times}$, so
\be
\boldsymbol{A} \, \boldsymbol{F_\textrm{w}} = 0 \, .
\label{AF0}
\ee
\noindent
We obtain the null streams $\boldsymbol{\widetilde{z}}$ by applying 
$\boldsymbol{A}$ to the network data vector:
\bea\label{eqn:nullstreammatrixform}
\boldsymbol{\widetilde{z}}
  & \equiv &  
         \boldsymbol{A} \, \boldsymbol{\widetilde{d}_\textrm{w}} \nonumber \\
  & = &  \boldsymbol{A} \, \boldsymbol{F_\textrm{w}} \, \boldsymbol{\widetilde{h}}
         + \boldsymbol{A} \, \boldsymbol{\widetilde{n}_\textrm{w}} \nonumber \\
  & = &  \boldsymbol{A} \, \boldsymbol{\widetilde{n}_\textrm{w}} \, .
\eea
\noindent
The two independent strain components $\tilde{h}_{+,\times}$ are
canceled out as a consequence of the definition \eqref{AF0} of the
null space, making each $\tilde{z}_\alpha$ a \emph{null stream}.

\begin{figure}
\resizebox{\columnwidth}{!}{\includegraphics{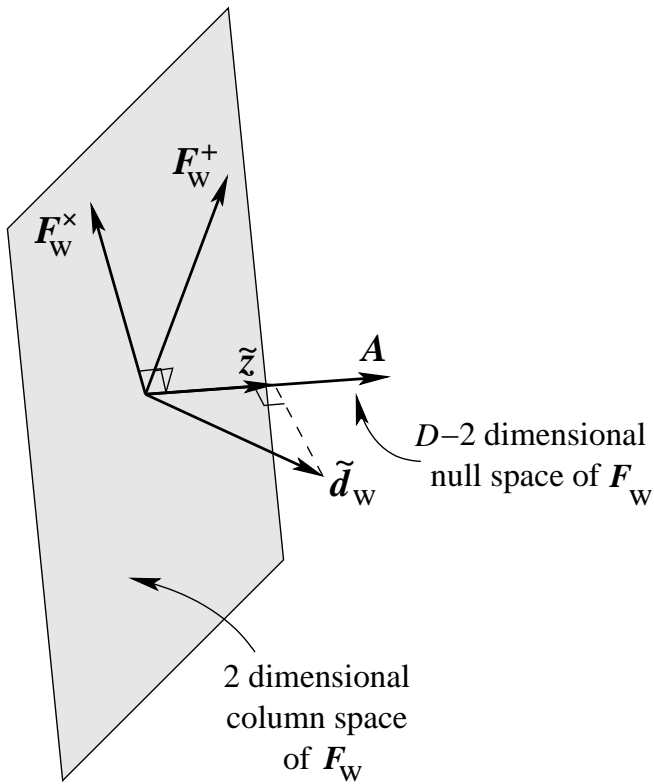}}
\caption{Geometry of the null stream construction for the 3-detector
  case. The null stream is obtained by projecting the data along the
vector $\boldsymbol{A}$, which is orthogonal to 
$\boldsymbol{F_{\textrm{w}}^+}$ and $\boldsymbol{F_{\textrm{w}}^\times}$.  
For $D$ non-aligned detectors $\boldsymbol{A}$ has $D-2$ dimensions.}
\label{fig:geometry}
\end{figure}

For example, the three-detector case is particularly simple.  In this
case the matrix $\boldsymbol{A}$ is equal to
\begin{equation}
\boldsymbol{A} =
\frac{\boldsymbol{F_\textrm{w}^+}\times\boldsymbol{F_\textrm{w}^\times}}
{|\boldsymbol{F_\textrm{w}^+}\times\boldsymbol{F_\textrm{w}^\times}|} \ .
\end{equation}
This is illustrated schematically in Figure~\ref{fig:geometry}.  In
the general case $\boldsymbol{A}$ can be obtained either via singular
value decomposition \cite{St:98} or by explicitly constructing the
associated projection operator (see Appendix~\ref{sec:projection} for
details).

Note that $\boldsymbol{A}$ is a function of the sky position (through
$F^{+,\times}_\alpha(\widehat{\Omega}_s)$) and frequency (through
$S_\alpha(f)$).  The GWB will only be canceled when $\boldsymbol{A}$
is evaluated for the correct source location, since generally
\be\label{eqn:wrongomega}
\boldsymbol{A}(\widehat{\Omega}, f) \, \boldsymbol{F_\textrm{w}}(\widehat{\Omega}', f) 
  \ne  0 \, .
\ee
\noindent
In the next subsection we will discuss how to deal with the case where
the source location is not known.

In the preceding discussion we have assumed implicitly that
$\boldsymbol{F_{\textrm{w}}^+}$ and $\boldsymbol{F_{\textrm{w}}^\times}$ are
independent.  In the general case the number of independent null
streams is $D-r$, where $r$ is the number of independent columns of
$\boldsymbol{F}$, i.e.
\be
r \equiv \mathrm{rank}(\boldsymbol{F_{\textrm{w}}}) \, .
\ee
\noindent
There are two cases
\begin{enumerate}
\item
  If at least one detector in the network has a different alignment
  from the others then $\boldsymbol{F_\textrm{w}^+}$ and 
  $\boldsymbol{F_\textrm{w}^\times}$ are independent and 
  $\mathrm{rank}(\boldsymbol{F_\textrm{w}})=2$.
  In this case there are $D-2$ null streams, and the method is
  applicable to networks of $D\ge3$ detectors.
\item
  If all detectors in the network are aligned then 
  $\boldsymbol{F_\textrm{w}^+} \propto \boldsymbol{F_\textrm{w}^\times}$ and
  $\mathrm{rank}(\boldsymbol{F_\textrm{w}})=1$.  In this case there are $D-1$
  null streams, and the method is applicable to networks of $D\ge2$
  detectors.
\end{enumerate}
Although in this paper we will concentrate on case 1, i.e. three or
more non-aligned interferometers, we emphasize that our method works
also with aligned detectors.

\subsection{Null-stream analysis}
\label{sec:analysis:nullstream2}

Since the $\boldsymbol{e}_i$ are orthonormal by
construction, it follows that
\be\n{eq:AAT}
\boldsymbol{A} \boldsymbol{A}^T = \boldsymbol{I}_{(D-2)\times(D-2)} \, .
\ee
\noindent
This implies that each $\tilde{z}_\alpha$ is a Gaussian random process
of unity variance and is uncorrelated with $\tilde{z}_\beta$, for all
$\beta \ne \alpha$.  This is the main advantage of the null-stream
formalism over other techniques: {\em the noise distribution of the
  projected network data is known\/} a priori, {\em regardless of the
  form of $h_{+,\times}$, under the assumption that a GWB from a
  particular direction is present}.  This allows us to perform
statistically significant tests of the network data, in particular to
test the hypothesis that a GWB from a given direction is present.

Since data from gravitational-wave detectors are sampled and digitized, 
in what follows we will consistently use discrete notation in our analysis 
of the statistics of the null stream.  Our conventions for discretely
sampled data are as follows: the Fourier-transform pair becomes
\bea
\tilde{x}[k]
& = &  \sum_{j=0}^{N-1} x[j] \,\e^{-i2\pi jk/N} \, , 
\nonumber \\
x[j] & = & \frac{1}{N} \sum_{k=0}^{N-1} \tilde{x}[k] \,\e^{i2\pi jk/N}
\, ,  
\eea 
where $N$ is the number of data points in the time domain.
Denoting the sampling rate by $f_s$, we can convert from continuous to
discrete notation using $x(t)\to x[j]$, $\tilde{x}(f)\to
f_s^{-1}\tilde{x}[k]$, $\int dt \to f_s^{-1}\sum_{j}$, $\int df \to
f_s N^{-1}\sum_{k}$, $\delta(t-t')\to f_s \delta_{jj'}$, and 
$\delta(f-f')\to N f_s^{-1}\delta_{kk'}$.  For example, the one-sided
strain noise power spectrum $S_\alpha[k]$ is 
\be\n{eq:discstrainnoise}
\langle 
    \tilde{n}_\alpha^{*}[k]
    \tilde{n}_\beta[k'] 
\rangle 
  =  \frac{N}{2} \delta_{\alpha\beta} \delta_{kk'} S_\alpha[k] \, .  
\ee 
\noindent
We will whiten the data by applying a zero-phase whitening filter
\cite{ChBlMaKa:04,chatterjiPhD}, and our normalization convention for
whitened data is
\be\n{eq:discwhitening}
\langle \tilde{n}_{\textrm{w}\alpha}^*[k]\tilde{n}_{\textrm{w}\beta}[k'] \rangle =
\delta_{\alpha\beta} \delta_{kk'} \, .  
\ee 
\noindent
The total energy in the null streams is 
\begin{equation}\n{eq:chi2}
E_{\textrm{null}} \equiv \sum_{\alpha=1}^{D-r} \sum_{k=0}^{N-1}
| \tilde{z}_\alpha[k]|^2 \, .
\end{equation}
Using \eqref{eq:AAT} and \eqref{eq:discwhitening} it follows that 
at the true source position $2E_{\textrm{null}}$ is $\chi^2$-distributed
with $2N(D-r)$ degrees of freedom.  In this case the expectation value 
of the null energy and its variance are both $N(D-r)$.

Although our considerations so far have assumed the sky position
$\widehat{\Omega}_s$ of the GWB source to be known {\em a priori}, in
practice this may not be the case.  Since we know that at the correct
source location the null energy will not contain any contribution from
the signal, a straightforward procedure is to scan over a grid of sky
positions in search of the minimum of the null energy.  
G\"ursel and Tinto \cite{GuTi:89} used time-delay estimates 
to limit their search to two possible regions of the sky (the 
regions around points $S$ and $S'$ in Figure~\ref{fig:ringgeometry}). 
This approach may fail when the duration of the
signals (or the timing uncertainty) is of the same order as the light
travel time between detectors, thus necessitating an all-sky search.
Since the
numerical analysis in \cite{GuTi:89} implies that the
characteristic angular width of the minimum of the null energy in a
neighborhood of the source location for a GWB 
with a central frequency of about $100$ Hz is equal to approximately 
$10^{-2}$ steradian for high SNRs, it follows that an all-sky search
should be performed over a sky grid containing more than $10^3$
resolvable directions.  
(Our numerical tests use a grid containing $10^4$ points.) 
In either case, for each trial direction
one postulates the presence of a gravitational-wave signal, forms a
linear combination of the detectors that is orthogonal to that
postulated direction, and $\chi^2$-tests this null stream for excess
energy.  If there exists a particular direction for which there is no
excess energy in the null stream, the data is regarded as consistent
with the hypothesis that a gravitational-wave burst is present and
incoming from the inferred direction. If, on the other hand,
$E_{\textrm{null}}$ is inconsistent with a $\chi^2$ distribution, then
one rejects the hypothesis that a GWB is present incoming from that
direction.  The best estimate of the source direction is taken as the 
direction with minimum $\chi^2$.

Although the null-stream method does not require knowledge of the two
GWB waveforms for its implementation, once the source location
$\widehat{\Omega}_s$ has been identified it is straightforward to
reconstruct $h_{+,\times}$ from the data themselves \cite{GuTi:89} (if
the detectors are all aligned then only one of the polarizations can
be reconstructed.)  The minimum-variance estimate of the two waveforms
for a network containing an arbitrary number of non-aligned detectors
is given in Appendix~\ref{sec:reconstruction}.

The null-stream combination of the data from a three-detector network
and the resulting $\chi^2$ test were first derived (in a different way)
by G\"ursel and Tinto \cite{GuTi:89} for detectors whose noises are
white.  They also implemented a near-optimal filtering procedure to
account for colored noise.  Our approach builds on this by
generalizing and simplifying the derivation of the null stream and
waveform reconstruction to networks containing an arbitrary number of
detectors with different colored noises.  One can also show that the 
null-stream procedure is formally equivalent to the maximum-likelihood 
analyses presented in \cite{FlHu:98b,Klimenko:05}, though we leave the 
demonstration to a future paper.

\subsection{Distinguishing GWBs from Noise Transients}

The numerical analysis performed by G\"ursel and Tinto in \cite{GuTi:89}
was not aimed at checking whether the null energy estimator could
distinguish GWBs from noise-induced glitches.  In fact, an analysis based
purely on the null stream runs into difficulty when applied to data
containing noise transients.  Strong uncorrelated glitches generally will not
cancel in the null stream combination because they are not correlated
in amplitude and phase in a way consistent with a GWB, implying a
$\chi^2>1$ per degree of freedom.  However, a null-stream analysis of
a real GWB may also produce $\chi^2>1$ per degree of freedom, due to
imperfect cancellation of the GWB in the null stream.  This may
happens for various reasons, such as the use of a discrete sky grid,
inaccurate calibration of the data, or imperfect whitening of
non-stationary data.  Thus, an analysis based purely on the null
stream would be forced to either reject both glitches and GWBs or
accept both. Further, counter to one's intuition, this problem could
get {\em worse} with stronger signals.

Glitches can also fool a null-stream analysis in 3-detector networks
when the transient is weak in at least one detector.  This is because
the non-observation of a signal by one detector $\alpha$ is {\em always} 
consistent with a GWB incident from a direction and
polarization to which the detector responses $F_{\textrm{w}\alpha}^+$ and
$F_{\textrm{w}\alpha}^\times$ are sufficiently small.  For sky positions with
$F_{\textrm{w}\alpha}^+ \sim F_{\textrm{w}\alpha}^\times \sim 0$ the null stream projection
matrix $\boldsymbol{A}$ reduces to
\be
\boldsymbol{A}_{\beta\gamma} 
  \to  \delta_{\alpha\beta}\delta_{\alpha\gamma} \, ,
       \qquad F_{\textrm{w}\alpha}^+, F_{\textrm{w}\alpha}^\times \to 0 \, .
\ee
\noindent
That is, the null stream for this sky position reduces to the detector in which there is no
transient.  This gives a $\chi^2$ per degree of freedom of order unity
regardless of whether the transient in the other two detectors is due
to a GWB or noise.  (Equivalently, when only two detectors observe a
signal, it is always possible to find $h_{+,\times}$ that fit the
output of these two detectors.)  Thus double-coincident glitches will
always pass a $\chi^2$ test for certain areas on the sky.  And while
networks containing four or more detectors will be less affected by
this problem \cite{Tinto96} because the size of the region of the sky
producing two or more simultaneous, below-threshold responses is
smaller than that for a single response, it will not be null.
 
In what follows we propose a simple way to make null-stream analyses
robust against glitches by comparing the amount of energy in the
null streams to that expected if the transients are uncorrelated.  
Let us consider how
the null stream energy $E_{\textrm{null}}$ depends on the individual
detector data streams $d_{\textrm{w}\alpha}$.  By defining the matrix
\be\n{eq:Q}
\boldsymbol{Q}  \equiv  \boldsymbol{A}^T \boldsymbol{A} \, ,
\ee
\noindent
we may write \eqref{eq:chi2} in the convenient form
\bea
\lefteqn{E_{\textrm{null}}(\widehat{\Omega}) 
    =    \sum_{k=0}^{N-1} \sum_{\alpha=1}^D \sum_{\beta=1}^D
         \tilde{d}^*_{\textrm{w}\alpha}[k] Q_{\alpha\beta}[k,\widehat{\Omega}] 
         \tilde{d}_{\textrm{w}\beta}[k] } \\
  & = &  \sum_{k=0}^{N-1}
\left[
\begin{array}{c c c c}
        \tilde{d}_{\textrm{w}1}^* &
        \! \hdots \!                    &
        \tilde{d}_{\textrm{w}D}^*
    \end{array}
\right] \!
\left[
    \begin{array}{c c c c}
        Q_{11} \! & \!  Q_{12} \! & \!  \hdots \! & \! Q_{1D} \\
        Q_{21} \! & \!  Q_{22} \! &               & \! Q_{2D} \\
        \vdots \! &               &               & \! \vdots \\
        Q_{D1} \! & \!  Q_{D2} \! & \!  \hdots \! & \! Q_{DD} \\
    \end{array}
\right] \!
\left[
    \begin{array}{c}
        \tilde{d}_{\textrm{w}1} \\
        \tilde{d}_{\textrm{w}2} \\
        \vdots                  \\
        \tilde{d}_{\textrm{w}D}
    \end{array}
\right] 
\, .  \nonumber \\
& & \label{eq:Qform}
\eea
\noindent
Note that the null energy contains contributions from both cross-correlation 
($\tilde{d}^*_{\textrm{w}\alpha}\tilde{d}^{\phantom{*}}_{\textrm{w}\beta}$) 
and auto-correlation 
($\tilde{d}^*_{\textrm{w}\alpha}\tilde{d}^{\phantom{*}}_{\textrm{w}\alpha}$) 
terms.  If the
signals in the various detectors are independent (as one might expect
for noise glitches), then the expectation value of the
cross-correlation terms in \eqref{eq:Qform} will be small 
compared to that of the auto-correlation terms.  In this case
the expectation value of $E_{\textrm{null}}$ is just the sum of the
diagonal terms in \eqref{eq:Qform}:
\be\label{eq:Qform1}
\textrm{mean}(E_{\textrm{null}}(\widehat{\Omega}))  \longrightarrow 
           \sum_{k=0}^{N - 1} \sum_{\alpha=1}^{D}
           Q_{\alpha\alpha}[k,\widehat{\Omega}]
           |\tilde{d}_{\textrm{w}\alpha}[k]|^2 
           \, . 
\ee
Here the mean is an ensemble average over noise instantiations.  This
observation motivates the use of a new energy measure, the 
{\em incoherent energy} $E_{\textrm{inc}}$, defined as the 
autocorrelation contribution to the null energy:
\be
E_{\textrm{inc}}(\widehat{\Omega}) 
  \equiv  \sum_{k=0}^{N - 1} \sum_{\alpha=1}^{D}  
          Q_{\alpha\alpha}[k,\widehat{\Omega}] 
          |\tilde{d}_{\textrm{w}\alpha}[k]|^2 
          \, .
\ee
If the transient signals in the various detectors are not correlated,
then we expect the following approximate equality to hold:
\be
\textrm{mean}(E_{\textrm{null}}) 
  \simeq  \textrm{mean}(E_{\textrm{inc}})  \, .
\ee
\noindent
If instead the signals in the detectors are correlated, as will be
the case when a GWB is present in the data, then at the correct sky
location the GWB contributions cancel in the null energy, and 
the following inequality should hold:
\be
\textrm{mean}(E_{\textrm{null}}) 
  <  \textrm{mean}(E_{\textrm{inc}})  \, .
\ee
\noindent
Thus, {\em a distinguishing feature of a GWB is that a significant
fraction of the energy in the individual detector data streams is
canceled in the null stream.\/}
 
In our simulations we test two simple measures of the degree 
to which $E_{\textrm{null}}$ and $E_{\textrm{inc}}$ show the 
behavior expected of a GWB. These are the quantities 
$E_{\textrm{null}} - E_{\textrm{inc}}$ and
$(E_{\textrm{null}}-E_{\textrm{inc}})/E_{\textrm{inc}}=E_{\textrm{null}}/E_{\textrm{inc}}-1$, which represent 
respectively the amount of ``correlated energy'' and the ratio 
of ``correlated energy'' to ``uncorrelated energy'' in the data.

An example is shown in Figure~\ref{fig:inc_vs_null}.  This figure 
shows the incoherent energy versus null energy for a GWB and a 
glitch of the same amplitude, evaluated for approximately $10^4$ 
uniformly distributed sky locations (see \ref{SECIII} for details).  
We note that for both the GWB and the glitch there are sky positions
for which the null energy is approximately unity, so that projection
onto the null energy alone does not distinguish this glitch and GWB.
However, glitches and GWBs scatter differently in terms of the two
energy measures $E_{\textrm{null}}$ and $E_{\textrm{inc}}$.
This suggests a modified procedure for distinguishing GWBs from
glitches: scan over the sky and look for directions for which 
$E_{\textrm{null}}$ is significantly smaller than $E_{\textrm{inc}}$.  
If there exists a sky direction for
which $E_{\textrm{null}}$ is sufficiently small compared to $E_{\textrm{inc}}$, 
we conclude that the transient could indeed be a GWB.  If instead
there is no direction for which $E_{\textrm{null}}$ is sufficiently 
small compared to $E_{\textrm{inc}}$, we
conclude that the transient is not a GWB.

Strong GWBs that are not precisely canceled in the null stream (due
to calibration errors, for example) and have $E_{\textrm{null}} >
N(D-r)$ will still pass this test because $E_{\textrm{null}} <
E_{\textrm{inc}}$.  Double-coincident glitches will fail even though
$E_{\textrm{null}} \simeq N(D-r)$ because they have $E_{\textrm{null}}
\simeq E_{\textrm{inc}}$.  Put another way, the incoherent energy
provides a natural cutoff in the significance of a $\chi^2$ per degree
of freedom measurement.  Intuitively, if the null stream cancels out
some large fraction of the excess incoherent energy, then we expect
the event to be a gravitational wave, even if $\chi^2>1$ per degree of
freedom.  The failure to cancel a significant portion of the incoherent 
energy will eliminate glitches even if $\chi^2 \simeq 1$ per degree of 
freedom.  

\setlength{\unitlength}{1in}
\begin{figure*}[tbp]
  \begin{center}
    \begin{picture}(6.5,2.5)
      \put(0,0){\rotatebox{0}{\resizebox{3in}{!}{\includegraphics{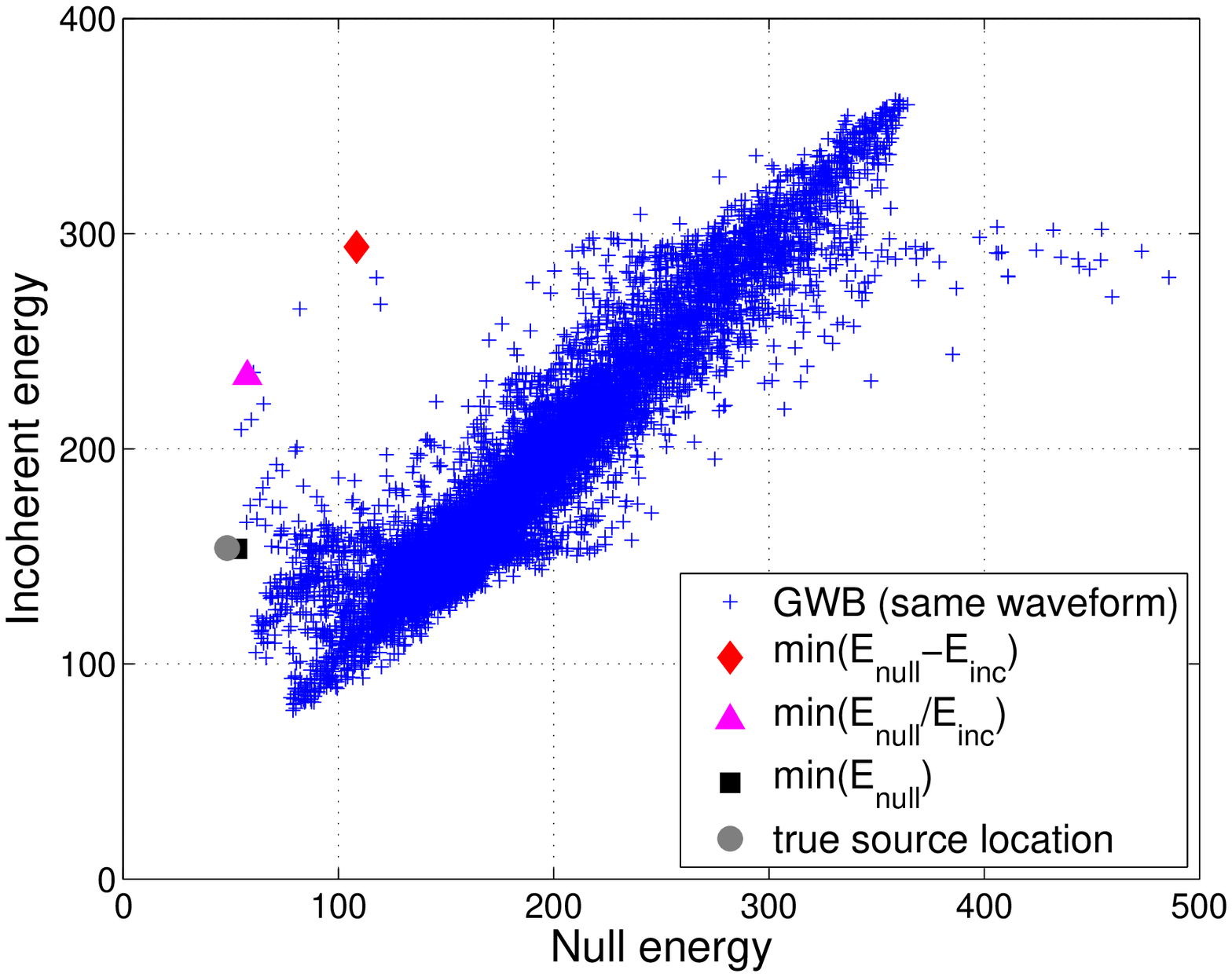}}}} 
      \put(3.5,0){\rotatebox{0}{\resizebox{3in}{!}{\includegraphics{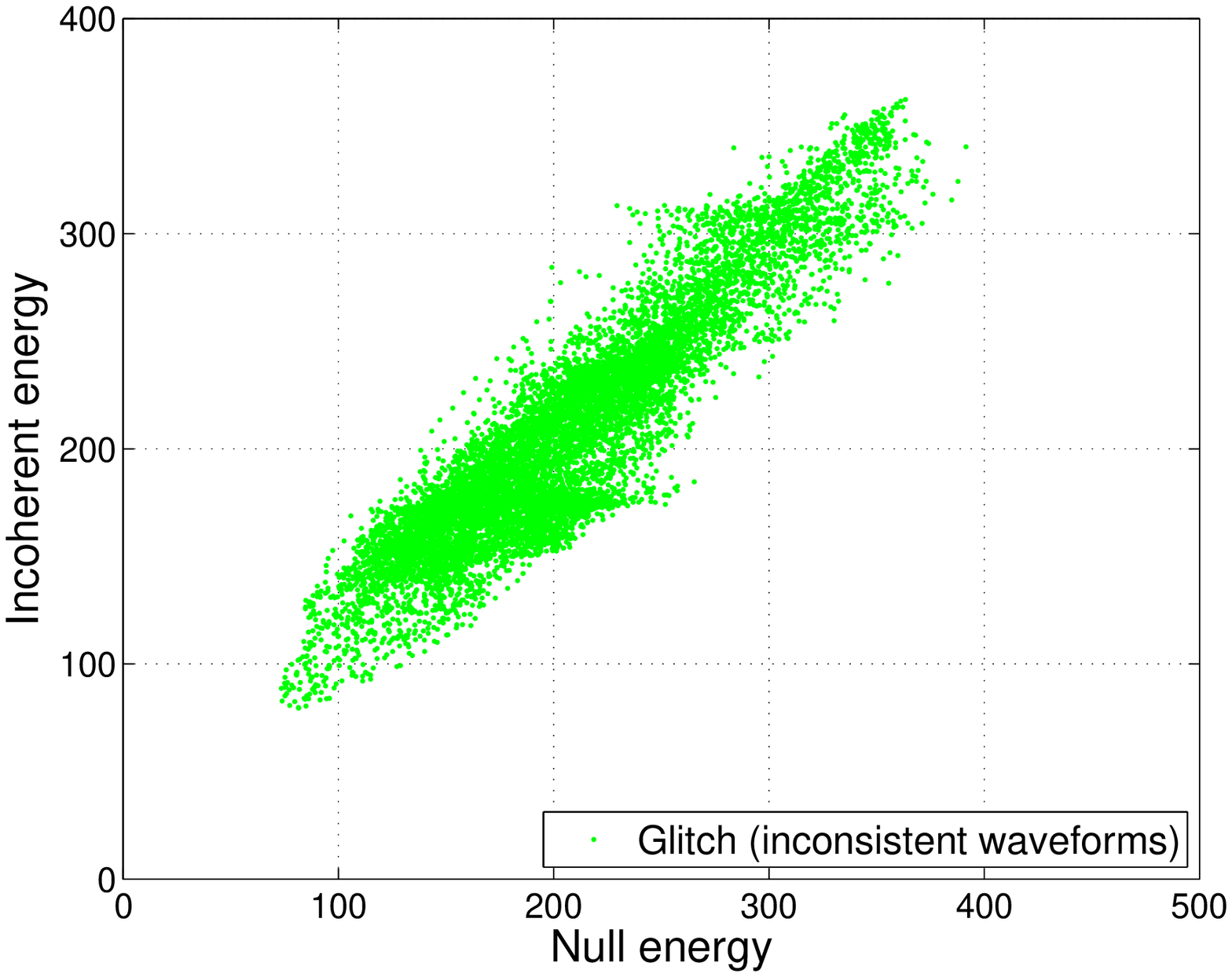}}}} 
    \end{picture}
  \caption{\label{fig:inc_vs_null}
    Scatter plot of the null and incoherent energies for a simulated
    GWB (left) and a simulated glitch (right) seen in a network
    consisting of the LIGO-Hanford and LIGO-Livingston 4 km detectors
    and Virgo.  In both cases the signal was scaled to rms SNR of 20 
    over the three detectors.  
    Each point represents one trial sky position;
    approximately $10^4$ sky positions in a uniform grid were tested.
    The waveforms and detector network used are discussed in detail in
    Section~\ref{SECIII}.  The GWB and glitch signals have the same
    signal-to-noise ratios and time delays in the individual
    detectors, and so are indistinguishable to incoherent tests.  Note
    that for both the GWB and the glitch there are sky positions for
    which the null energy is consistent with noise ($E_{\textrm{null}}
    \simeq N(D-r) = 60$ for these simulations).  However, for the GWB there are also sky
    positions with $E_{\textrm{null}} \ll E_{\textrm{inc}}$ (points
    above the diagonal); these are due to the fact that the GWB signal
    is correlated between the detectors.  The glitch signal does not
    access this portion of the $(E_{\textrm{null}},E_{\textrm{inc}})$
    space.  This observation is the basis of our consistency test; we
    scan over the sky and look for directions where $E_{\textrm{null}}
    < E_{\textrm{inc}}$. The true source location is indicated by the
    circle on the GWB plot.  Also shown on the GWB plot are the tested
    sky positions which have the smallest values of
    $E_{\textrm{null}}-E_{\textrm{inc}}$,
    $E_{\textrm{null}}/E_{\textrm{inc}}$, and $E_{\textrm{null}}$.  The
    sky maps for these same simulations are shown in
    Figure~\ref{fig:skymaps}.}
\end{center}
\end{figure*}

To get more insight into the behavior and usefulness of these energy
measures, let us consider in more detail their expectation values over
noise instantiations.  Allowing for the case in which the transient
is not a GWB, we write
\be\n{eq:dgeneral}
\tilde{d}_{\textrm{w}\alpha}[k]
  =  \tilde{n}_{\textrm{w}\alpha}[k] + \tilde{g}_{\textrm{w}\alpha}[k] \, ,
\ee
\noindent
\noindent
where $g_{\textrm{w}\alpha}$ denotes the noise-weighted transient as seen in
detector $\alpha$.  For example, if the transient {\em is} a GWB from
the direction $\widehat{\Omega}_s$, and the tested sky position is
$\widehat{\Omega}$, then
\bea\n{eq:g}
\tilde{g}_{\textrm{w}\alpha}[k] 
 &=& \frac{
         F^+_{\alpha}[k,\widehat{\Omega}_s] \, \tilde{h}_+[k]
         + F^\times_{\alpha}[k,\widehat{\Omega}_s] \, \tilde{h}_\times[k]
     }{
         \sqrt{\frac{N}{2} S_\alpha[k]}
     }
     \mbox{} \nonumber \\
 & & \mbox{} \times \e^{i2\pi f_s k/N (
         \Delta t_\alpha(\widehat{\Omega})
         -\Delta t_\alpha(\widehat{\Omega}_s)
     )} \, .
\eea
\noindent
The phase term
accounts for time shifting based on the incorrect sky location
$\widehat{\Omega}$ instead of the true but unknown sky location
$\widehat{\Omega}_s$.

For simplicity, let us restrict ourselves to the case of detectors with 
equal noise spectra, for which $\boldsymbol{Q}$ is independent of 
frequency.  
Then the noise-weighted transient signal $g_{\textrm{w}\alpha}$ appears in the 
null and incoherent energies in the combination
\bea
\rho^2_{\alpha\beta}
  & \equiv &  \sum_{k=0}^{N-1} \tilde{g}^*_{\textrm{w}\alpha}[k]\tilde{g}^{\phantom{*}}_{\textrm{w}\beta}[k] 
    =  \sum_{k=0}^{N-1} \frac{
        \tilde{g}^{*}_\alpha[k]\tilde{g}^{\phantom{*}}_\beta[k] 
    }{
        \frac{N}{2} \sqrt{S_\alpha[k]S_\beta[k]}
    }
    \n{eqn:crossSNR} \, , ~\mbox{}
\eea
which is the noise-weighted cross-correlation of the signals in detectors 
$\alpha$ and $\beta$.  The diagonal terms are
\bea
\rho^2_{\alpha\alpha}
  & \equiv &  
    \sum_{k=0}^{N-1} |\tilde{g}_{\textrm{w}\alpha}[k]|^2
    =  \sum_{k=0}^{N-1} \frac{
        |\tilde{g}_\alpha[k]|^2
    }{ 
        \frac{N}{2} S_\alpha[k]
    } \n{eqn:SNR} \\
  & \leftrightarrow &
    2\int_{-\infty}^{\infty} \!\!\! df 
    \frac{|\tilde{g}_\alpha(f)|^2 }{S_\alpha(f)}  
    \, . \nonumber 
\eea
This autocorrelation term is
simply the squared signal-to-noise ratio of an optimal matched filter
for the transient in detector $\alpha$, as indicated by the second line 
of \eqref{eqn:SNR}.  

As an example, let us consider the special case of a linearly
polarized GWB (e.g., with $h_\times=0$), 
with trial sky position $\widehat{\Omega}$ and true 
source position $\widehat{\Omega}_s$. In this case the cross-SNR
is equal to
\bea\n{eqn:crossSNRlinearcase}
\rho^2_{\alpha\beta}  & \longrightarrow &  
    \frac{2}{N} \sum_{k=0}^{N-1} 
    F^+_\alpha[k,\widehat{\Omega}_s] F^+_\beta[k,\widehat{\Omega}_s] 
    \, \times \nonumber \\
    & & \times \,
    \frac{|\tilde{h}_+[k]|^2}{\sqrt{S_\alpha[k]S_\beta[k]}} 
    \cos{\Psi_{\alpha\beta}(\widehat{\Omega},\widehat{\Omega}_s)} 
\eea
(only the real part of $\rho^2_{\alpha\beta}$ contributes to the energies), 
where the phase error is 
\bea\n{eqn:phaseerror}
\Psi_{\alpha\beta}(\widehat{\Omega},\widehat{\Omega}_s)
 &=& 2\pi \frac{f_s k}{N} (
         \Delta t_\alpha(\widehat{\Omega})
         -\Delta t_\alpha(\widehat{\Omega}_s)
    \mbox{} \nonumber \\
 & & \mbox{}
         -\Delta t_\beta(\widehat{\Omega})
         +\Delta t_\beta(\widehat{\Omega}_s)
     ) \, .
\eea
\noindent
The cross-SNR $\rho^2_{\alpha\beta}$ for $\alpha\ne\beta$ is typically 
of the same order of magnitude
as $\rho^2_{\alpha\alpha}$, but it is positive or negative depending on the
timing error between pairs of detectors.  As we test different
positions on the sky, the timing errors change and the GWB
contributions from the different detectors move in and out of phase.
This will produce interference fringes in maps of $E_{\textrm{null}}$, 
$E_{\textrm{null}}-E_{\textrm{inc}}$, and
$E_{\textrm{null}}/E_{\textrm{inc}}$.  The location and spacing of these
fringes are determined by the dominant frequency of the signal, the
true sky position of the source, and the detector geometries and
relative locations.

In terms of the SNRs \eqref{eqn:crossSNR}, \eqref{eqn:SNR}, 
the expectation values of the lowest moments of $E_{\textrm{null}}$, 
$E_{\textrm{inc}}$, and $E_{\textrm{null}}-E_{\textrm{inc}}$ are
\bea\n{eqn:Moments_n}
\textrm{mean}(E_{\textrm{null}})
  & = &  N(D-r) + \sum_{\alpha=1}^D \sum_{\beta=1}^D 
         Q_{\alpha\beta}\rho^2_{\alpha\beta}  \, , \\
\textrm{var}(E_{\textrm{null}})
  & = &  N(D-r) + 2 \sum_{\alpha=1}^D \sum_{\beta=1}^D 
         Q_{\alpha\beta}\rho^2_{\alpha\beta}  \, , 
\eea
\bea\n{eqn:Moments_i}
\textrm{mean}(E_{\textrm{inc}})
  & = &  N(D-r) + \sum_{\alpha=1}^D Q_{\alpha\alpha}\rho^2_{\alpha\alpha} \, ,\\
\textrm{var}(E_{\textrm{inc}})
  & = &  N\sum_{\alpha=1}^D Q_{\alpha\alpha}^2 
         + 2\sum_{\alpha=1}^D Q_{\alpha\alpha}^2\rho^2_{\alpha\alpha}  \, ,
\eea
\bea
\textrm{mean}(E_{\textrm{null}}-E_{\textrm{inc}})
  & = &  \sum_{\alpha=1}^D \sum_{\beta=1}^D (1-\delta_{\alpha\beta}) 
         Q_{\alpha\beta}\rho^2_{\alpha\beta}  \, , \\
\textrm{var}(E_{\textrm{null}}-E_{\textrm{inc}})
  & = &  N\sum_{\alpha=1}^D \sum_{\beta=1}^D (1-\delta_{\alpha\beta}) 
         Q_{\alpha\beta}^2 
         + \mbox{} \nonumber \\
  &   &  \hspace{-1.25in} \mbox{} + 2\sum_{\alpha=1}^D \sum_{\beta=1}^D \sum_{\gamma=1}^D 
         (1-\delta_{\alpha\beta}) (1-\delta_{\alpha\gamma}) 
         Q_{\alpha\beta} Q_{\alpha\gamma} \rho_{\alpha\gamma}^2 \, . \quad
\label{eqn:Moments_ni}
\eea
\noindent
We note that the signal enters $E_{\textrm{inc}}$ only through its SNR
$\rho^2_{\alpha\alpha}$ in the individual detectors; $E_{\textrm{inc}}$ does
not depend on the structure of the transient signal.  As a result,
variations in $E_{\textrm{inc}}$ reflect variations in the network
sensitivity due to noise-weighted geometrical factors (the
$\boldsymbol{Q}$) and do not contain significant information on the
signal.  By contrast, $E_{\textrm{null}}-E_{\textrm{inc}}$ contains 
only cross terms, and shows the
interference of the signals measured by the different detectors.

For example, Figure~\ref{fig:skymaps} shows sky maps of the
$E_{\textrm{null}}$, $E_{\textrm{inc}}$, and
$E_{\textrm{null}}/E_{\textrm{inc}}$ [note that 
$E_{\textrm{null}}/E_{\textrm{inc}}=(E_{\textrm{null}}-E_{\textrm{inc}})/E_{\textrm{inc}}+1$] 
for the same GWB and glitch
signals used in Figure~\ref{fig:inc_vs_null}.  The null energy maps
for the GWB and the glitch are very similar.  
The GWB and glitch are constructed to have
the same relative time delays and SNRs in the various detectors.  As a
result, the incoherent energy maps are virtually identical for the GWB
and the glitch.  Removing this signal-independent structure from the
null energy makes the signal-dependent structure in the sky maps
clearer.  In particular, the plot of
$E_{\textrm{null}}/E_{\textrm{inc}}$ for the GWB shows sharp
interference fringes orthogonal to the Hanford-Livingston and 
Livingston-Virgo baselines (the signal was strongest in Livingston 
and Virgo in this simulation).  The sky location
of the GWB signal lies on one of the two intersection points of these
interference fringes, so they can be used to locate the source.
Such sharp features are not present in the corresponding sky map for
the glitch, since the glitch waveforms are not strongly correlated.

\setlength{\unitlength}{1in}
\begin{figure*}[tbp]
  \begin{center}
    \begin{picture}(6,7.5)
      \put(0,5){\rotatebox{0}{\resizebox{3in}{!}{\includegraphics{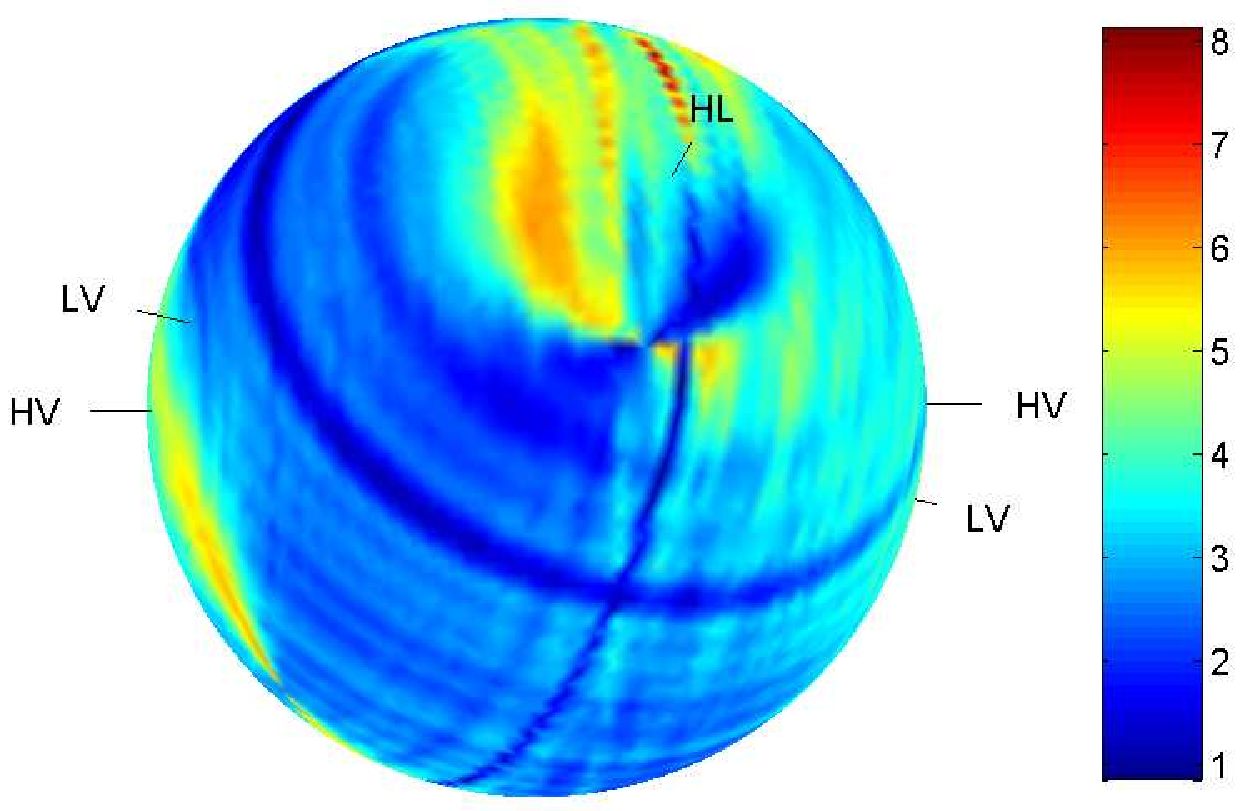}}}} 
      \put(0,2.5){\rotatebox{0}{\resizebox{3in}{!}{\includegraphics{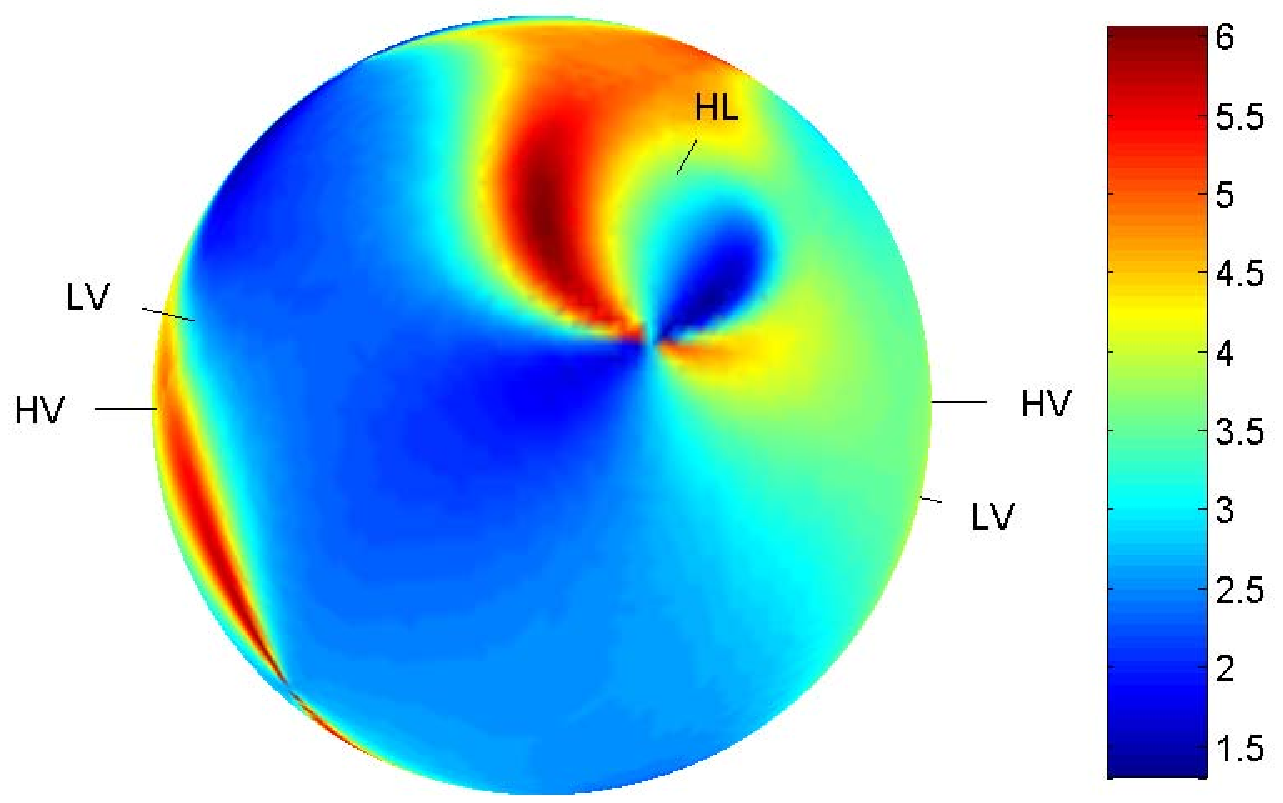}}}} 
      \put(0,0){\rotatebox{0}{\resizebox{3in}{!}{\includegraphics{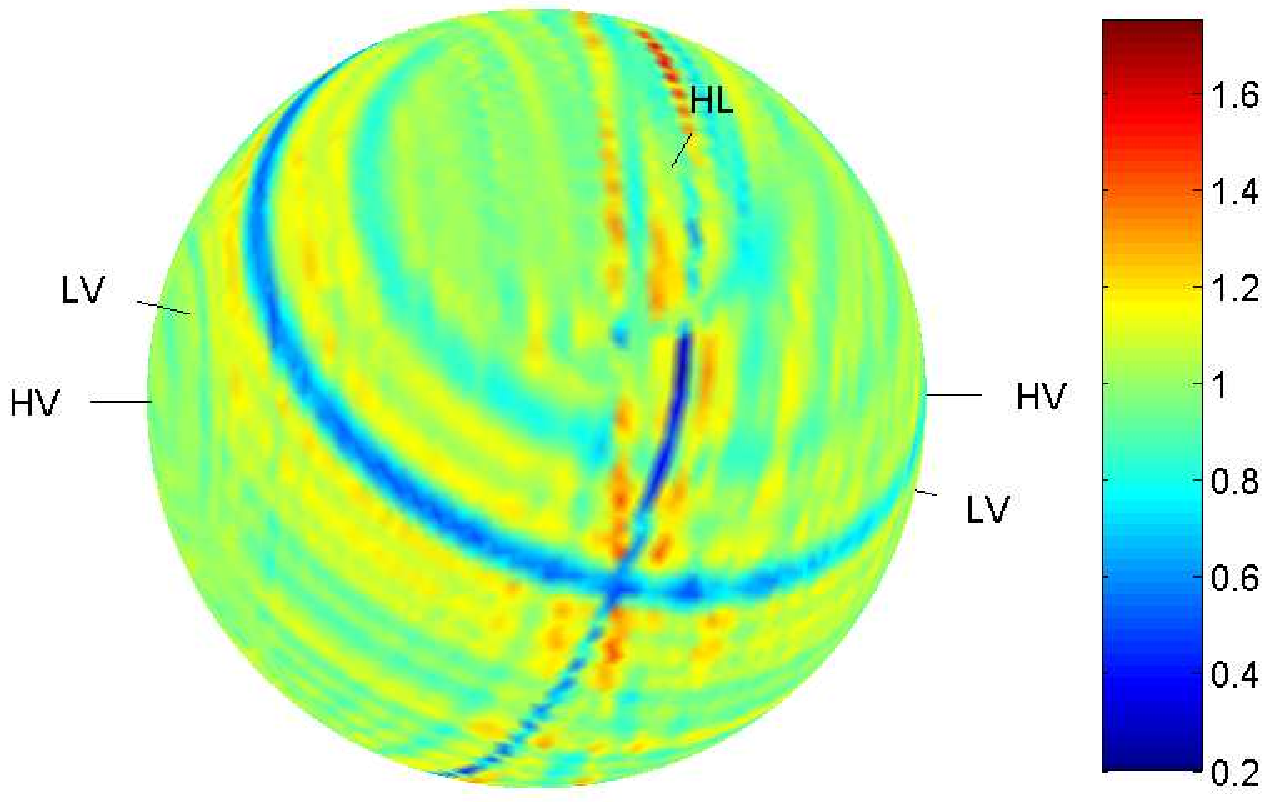}}}} 
      \put(3,5){\rotatebox{0}{\resizebox{3in}{!}{\includegraphics{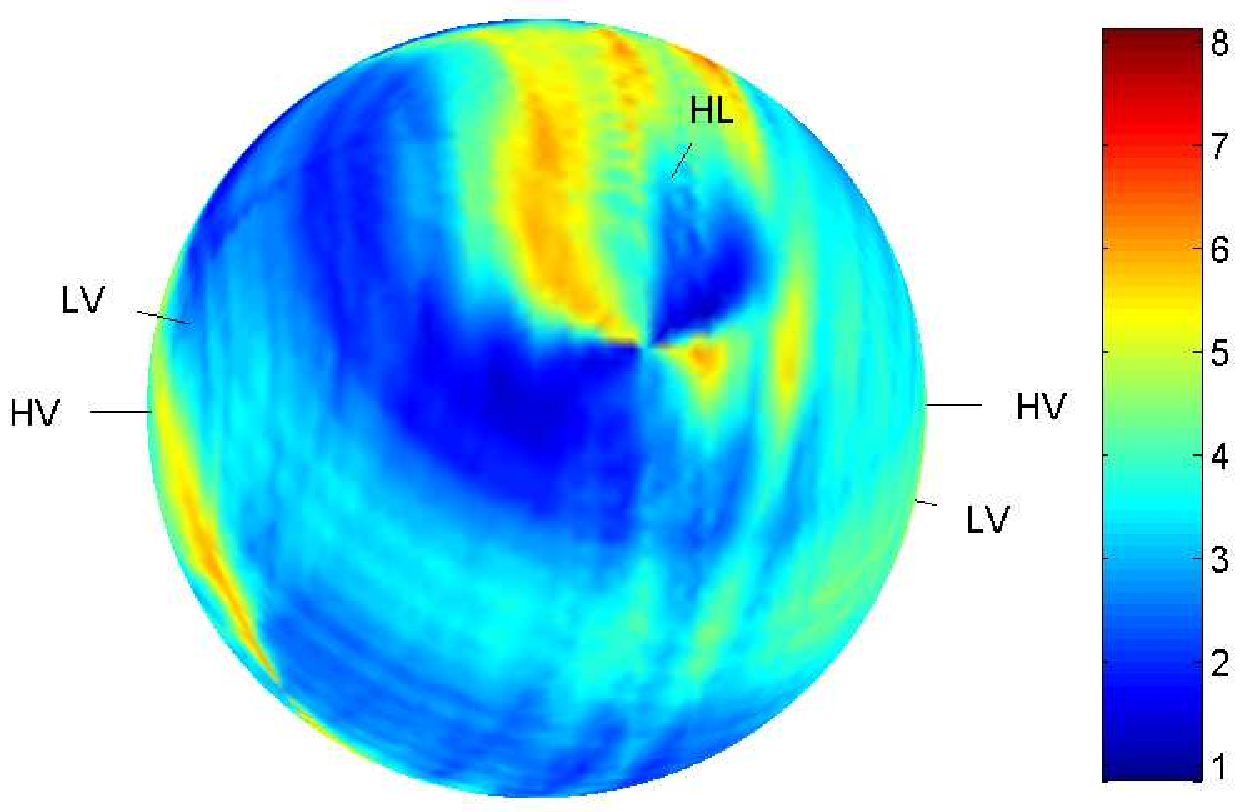}}}} 
      \put(3,2.5){\rotatebox{0}{\resizebox{3in}{!}{\includegraphics{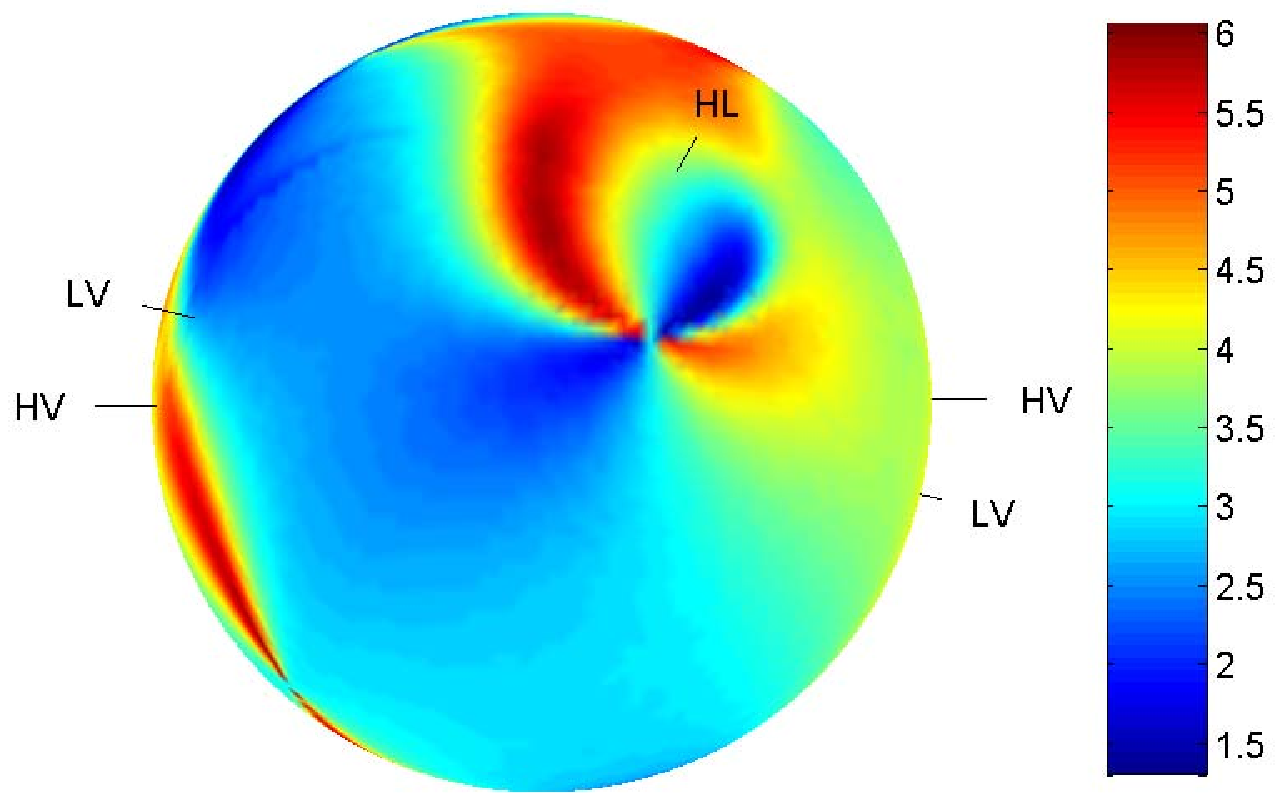}}}} 
      \put(3,0){\rotatebox{0}{\resizebox{3in}{!}{\includegraphics{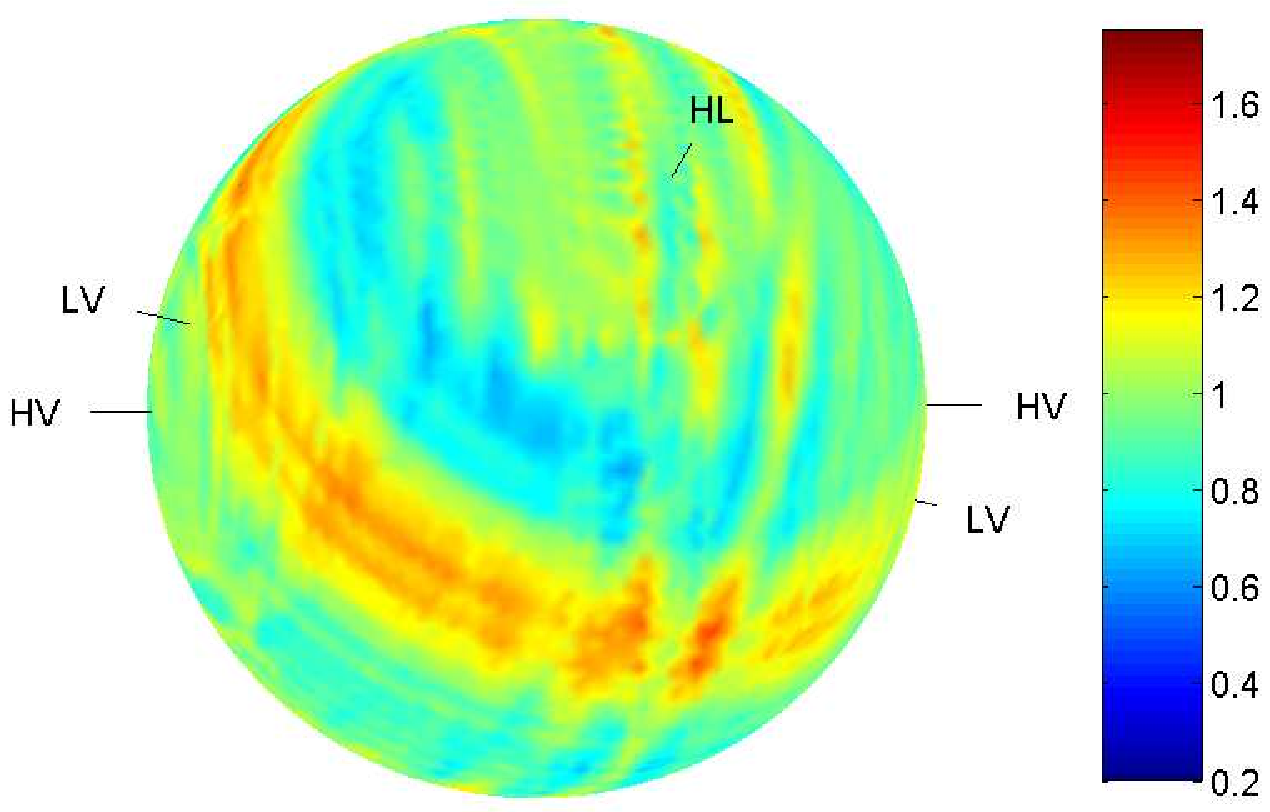}}}} 
      \put(0,7.05){GWB: $E_{\mathrm{null}}$}
      \put(0,4.65){GWB: $E_{\mathrm{inc}}$}
      \put(0,2.15){GWB: $E_{\mathrm{null}}/E_{\mathrm{inc}}$}
      \put(3,7.05){glitch: $E_{\mathrm{null}}$}
      \put(3,4.65){glitch: $E_{\mathrm{inc}}$}
      \put(3,2.15){glitch: $E_{\mathrm{null}}/E_{\mathrm{inc}}$}
    \end{picture}
\caption{\label{fig:skymaps}
  Sample sky maps, normalized to $E/(N(D-r))$, for the same GWB and
  glitch signals used in Figure~\ref{fig:inc_vs_null}.  The plots on
  the left are for a GWB with rms SNR of $20$ in the three detectors.
  Those on the right are for a glitch with the same relative time
  delays and signal energies in each detector as the GWB.  (Note that
  these two events are indistinguishable to an incoherent analysis.)
  The network consists of the LIGO-Hanford (H) and LIGO-Livingston (L) 4 km
  detectors and Virgo (V).  The null energy map (top) shows interference
  fringes due to the transient signal, as well as structure due to the
  network geometry.  They are very similar for the GWB and the glitch.
  The incoherent energy map (middle) is constructed from the
  auto-correlations of the individual detector data streams and
  reflects variations in the network sensitivity over the sky. It is
  virtually identical for the GWB and the glitch because the two
  events have the same relative time delays and SNRs.  Dividing out the 
  incoherent energy from the null energy (bottom) removes this
  structure associated with the network geometry, making the 
  signal-dependent structure clearer. The GWB map shows sharp
  interference fringes (blue and red rings) where the time delays 
  along the H-L, L-V, and H-V baselines match those of the source 
  location. There is little of such structure in the glitch map.}
\end{center}
\end{figure*}

We note from \eqref{eqn:Moments_n}-\eqref{eqn:Moments_ni} that fractional 
fluctuations in the energies scale as $N^{-1/2}$ for weak
signals.  Since the signal-dependent structure scales as
$\rho^2_{\alpha\beta}$, the basic limit of purely local measurements
of sky maps scales as $\rho^2_{\alpha\beta}/N^{1/2}$.  This is the same scaling as
in excess-power searches \cite{AnBrCrFl:01}.

Also, we note that if we restrict our analysis to a frequency range
$f = k f_s / N \in[f_{\textrm{min}},f_{\textrm{max}}]$, then $N$ in
\eqref{eqn:Moments_n}-\eqref{eqn:Moments_ni} becomes the number of
frequency bins actually summed over, and $\rho^2$ \eqref{eqn:crossSNR},
\eqref{eqn:SNR} is to be computed over the same (positive) frequency
range.

Finally, we point out that in the high SNR limit 
the minimum of the null energy $E_{\textrm{null}}$
occurs at the source location \cite{GuTi:89}, which is
useful for solving the inverse problem for bursts.  
The incoherent energy $E_{\textrm{inc}}$ is {\em not} an extremum 
at the source location, so measures like $E_{\textrm{null}}/E_{\textrm{inc}}$ 
and $E_{\textrm{null}}-E_{\textrm{inc}}$ cannot directly resolve the source
location.  As noted above, however, the clearer signal-dependent 
structure in $E_{\textrm{null}}-E_{\textrm{inc}}$ and 
$E_{\textrm{null}}/E_{\textrm{inc}}$ compared to $E_{\textrm{null}}$ 
may be useful for this purpose.  In any case, our consistency test can 
improve the detection confidence of a gravitational-wave burst, and
should be regarded as an essential first step in solving the inverse
problem.

\subsection{Sensitivity of the method to data calibration}
\label{sec:calibration}

The derivation of the null-stream combinations described in subsection
\ref{sec:analysis:nullstream} assumed the data from the
interferometers to be perfectly calibrated. This means that properly
modeled transfer functions of the interferometer responses (which
depend on various parameters associated with the characteristics of
the cavities of the interferometers) are applied to the
raw data in order to obtain strain measurements. In practice, however,
the parameters describing these transfer functions are
known with finite accuracy, in turn preventing the null-stream
combinations from exactly canceling the GWB signal at the correct
source location. In order to quantify 
the effect on our method we should note that the magnitude of the
residual signal in the null-stream combinations will be proportional
to the accuracy by which the calibration parameters are known. 
For the LIGO detectors the calibration parameters are known 
to better than ten percent \cite{Calibration}.
Preliminary studies indicate our consistency test is robust 
against calibration errors of this size; these effects will 
be studied in more detail in a future article.

\subsection{Implementation}
\label{sec:implementation}

Our null-stream based analysis has been implemented as a publicly
available \textsc{Matlab} package, ``\textsc{Xpipeline}''
\cite{Xpipeline}.  For a specific detector network and event time,
\textsc{Xpipeline} reads the appropriate data from frame files
\cite{frames} (the standard format for storing data from
gravitational-wave detectors), optionally injects GWB signals and/or
glitches, whitens the data, computes the null stream coefficient
matrix $\boldsymbol{A}$ for each specified sky direction and
frequency, computes the time shifts for each direction, steps through
data in overlapping blocks of user-specified duration, time shifts the
data to the nearest sample, Fourier transforms it, completes the time
shift with a phase rotation, forms the null stream in the frequency
domain, sums the power in user-specified frequency bands, and records
the null and incoherent energies for each time-frequency band and
direction.  \textsc{Xpipeline} runs in approximately 1/100th real
time.  For example, the analysis of the $10^4$ simulated events used
to produce Figures~\ref{fig:scatter_null_over_inc}-\ref{fig:ROC_null}
took approximately $16$ hours on $4$ Intel Pentium 4, $2.66$ GHz
computers.  This makes our null-stream-based consistency
test feasible as a follow-up test in GWB searches.

\section{Simulations}
\label{SECIII}

\subsection{Network and signal types}

To test the efficacy of our statistical test in discriminating GWBs
from noise glitches in the $(E_{\textrm{null}}, E_{\textrm{inc}})$ space, we
need to select a detector network, a population of GWBs, and a
population of glitches.

We elect to simulate a network consisting of the Hanford and
Livingston 4~km interferometers (``H'' and ``L'') and a third
identical instrument at the Virgo site (``V''). For the sake of
simplicity we neglect both the Hanford 2~km interferometer and the
differences between the LIGO and Virgo design sensitivities.  The
locations $\vec{r}_\alpha$ and orientations (which determine $F^+$,
$F^\times$) of the interferometers are taken from
\cite{Al_etal:01,AnBrCrFl:01}.  The calibrated stationary background
noise $n_\alpha[j]$ for H and L are taken from a
standard 24 hour reference simulation \cite{LIGO-VIRGO-data}.  The
background noise for V is taken from the Hanford simulation with a 2
second time shift, which is much larger than the time scales of the
signals used in this analysis.

The next step is to select the GWB and glitch waveforms.  To simulate
a glitch, we need three waveforms (one for each detector) that are not
strongly correlated.  To simulate GWBs we wish to use waveforms that
are motivated by astrophysical considerations.  Our consistency test
is ultimately based on the fact that for a GWB the signal seen in each
detector is correlated in a particular way, whereas for a glitch the
signals generally will not be correlated.  In order to show that our
test does not rely on any fundamental difference between ``typical''
GWB waveforms and ``typical'' glitch waveforms, we elect to use the
{\em same} set of waveforms for simulating GWBs and glitches.  We
select three representative waveforms from the
Dimmelmeier-Font-Mueller (``DFM'') catalog \cite{DiFoMu:02b} of Type II
core-collapse supernovae.  Specifically, we choose the A1B3G3
``regular collapse'' waveform, the A1B3G5 ``rapid collapse'' waveform,
and the A3B4G2 ``multiple bounce'' waveform.
Figure~\ref{fig:DFMwaveforms} shows the time-series and power spectra
of these waveforms.  As we shall see below, these three waveforms have
moderately low but nonzero cross-correlations.

\setlength{\unitlength}{1in}
\begin{figure*}[tbp]
  \begin{center}
    \begin{picture}(6.5,2.5)
      \put(0,0){\rotatebox{0}{\resizebox{3in}{!}{\includegraphics{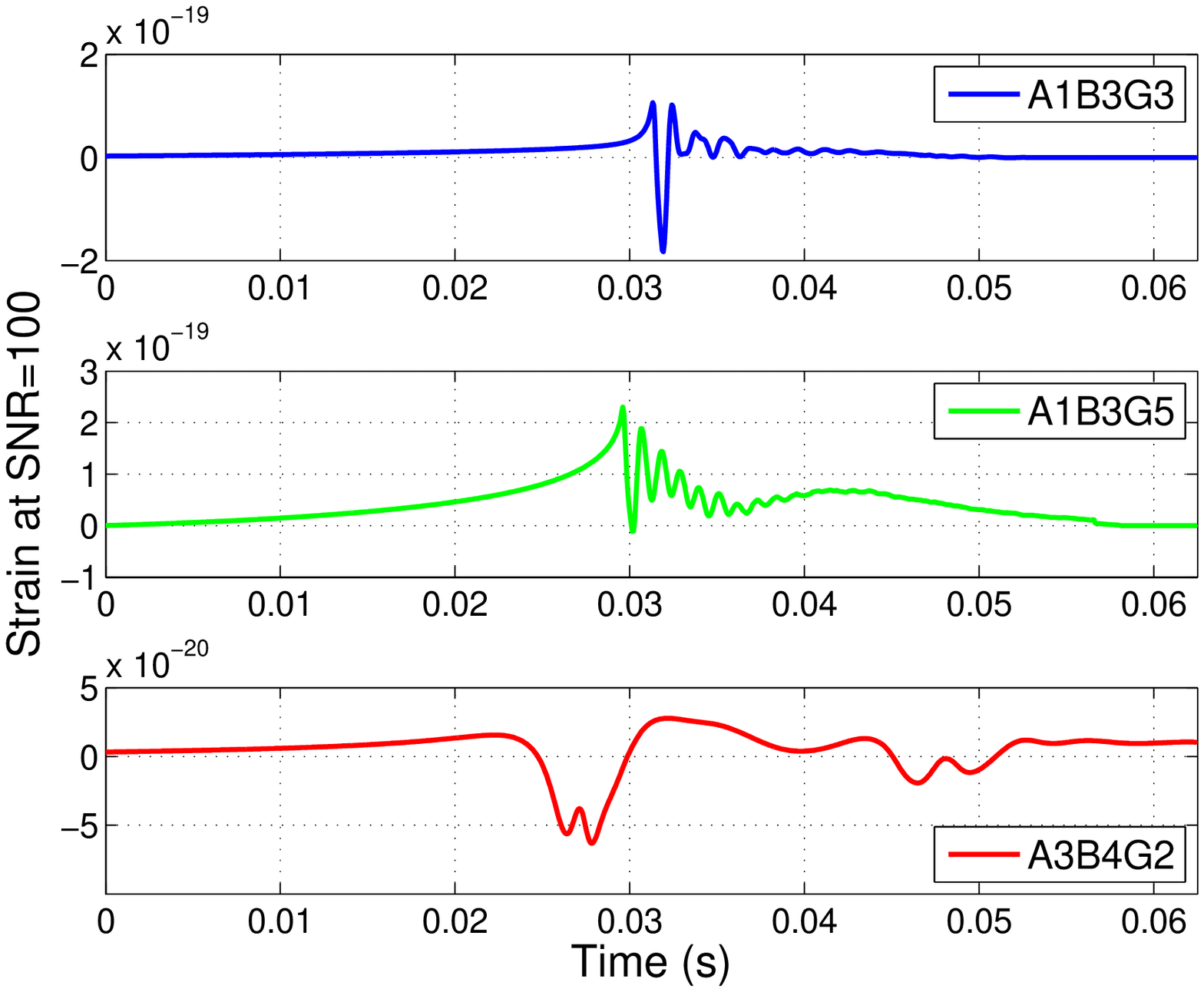}}}} 
      \put(3.5,0){\rotatebox{0}{\resizebox{3in}{!}{\includegraphics{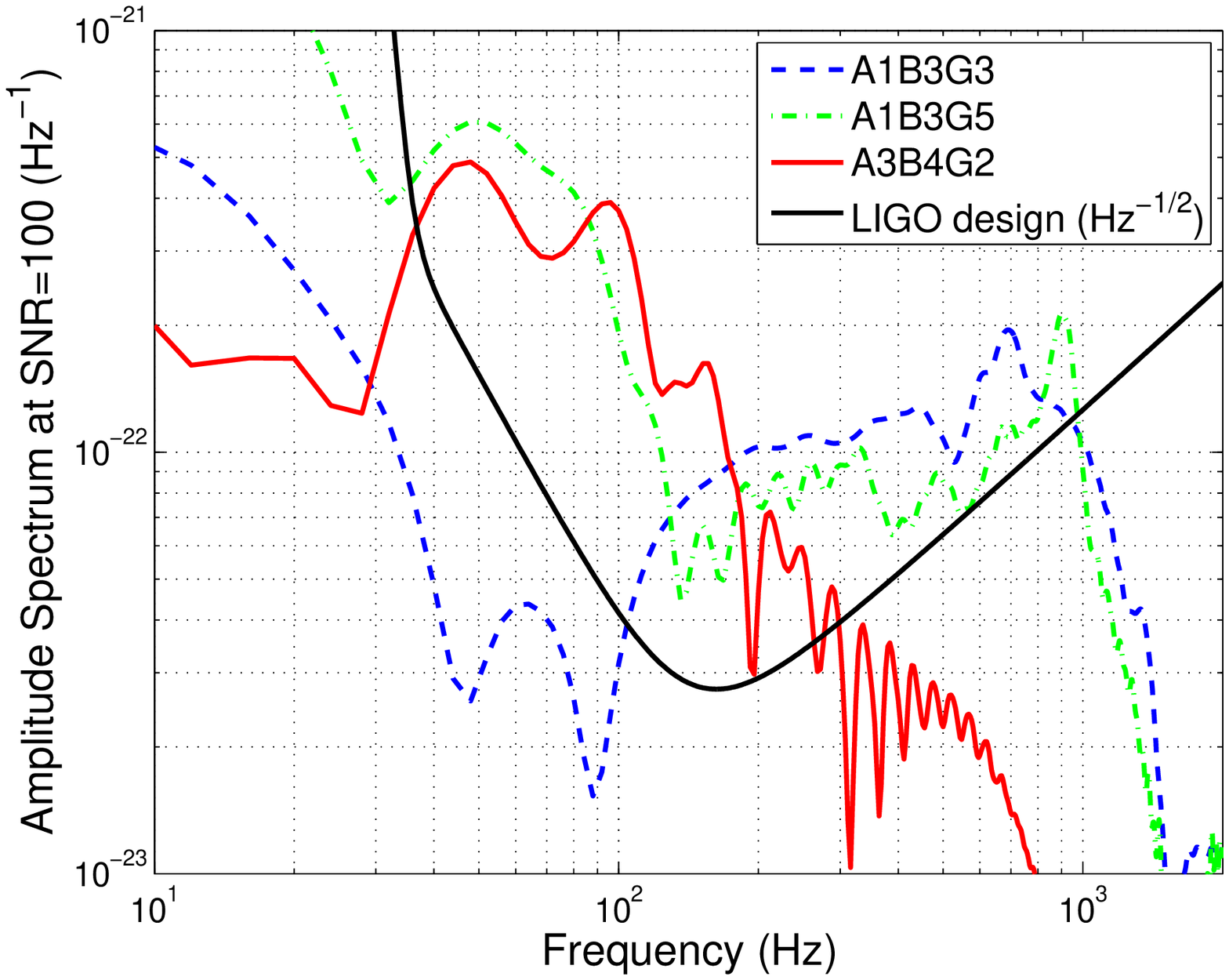}}}} 
    \end{picture}
  \caption{\label{fig:DFMwaveforms}
    Dimmelmeier - Font - Mueller (``DFM'') waveforms used to simulate
    GWBs and glitches in our analysis. Left plot: time series
    waveforms scaled to SNR = 100 for optimal orientation. (The linear
    trend present in the original waveforms has been removed.) Right
    plot: spectra with a $1$ Hz resolution; shown for comparison is
    the LIGO design noise curve used for both LIGO and Virgo
    detectors.  }
\end{center}
\end{figure*}

To simulate a gravitational-wave signal, one of the three waveforms is
randomly selected and added into the data stream from each
interferometer, with time delays and scaling appropriate for some
choice of polarization angle and location on the sky.  To simulate a
glitch we follow the same procedure, except that a different waveform
is selected for each detector.  The scaling and time delays proceed
as for a GWB.  This population of glitches has the property that it
would pass any incoherent test, as the arrival times, power
distribution, and even frequency bands are consistent with those of
true gravitational-wave signals.  We stress that, though not a
realistic glitch population, this provides us with examples of the
kind of pathological glitches that cannot be dismissed by per-detector
methods.

\subsection{Analysis parameters}

The main tunable parameters in our analysis are the time and frequency
bands and the sky positions to test.  Ideally the integration time and
frequency band should be matched to the signal being tested, to
maximize the signal-to-noise ratios $\rho^2_{\alpha \alpha}$ and
minimize the background noise contribution $N(D-r)$.  We choose an
integration length of $1/16$ s, which is the smallest power of $2$
larger than the durations of the three sample waveforms tested.  We
overlap consecutive data segments by 50\% to minimize the loss of
signal-to-noise when a signal overlaps the edge of a data segment.  We
use a single frequency band \footnote{ \textsc{Xpipeline} can also
  process multiple frequency bands; we use a single band in this
  demonstration for simplicity.  } of [64,1024]Hz.  The upper
frequency limit is set by the highest frequency at which our target
signals have significant power compared to the noise curve; see
Figure~\ref{fig:DFMwaveforms}.  The lower limit is set at 64 Hz
because the actual noise level in current detectors is larger than the
design noise below this cutoff \cite{LIGOnoise}.

Once the locations of the detectors and the upper
bound on frequencies involved in the analysis are
known, a set of directions $\{\widehat{\Omega}\}$ covering the sky
may be produced.  Both the projection matrix $\mathbf{Q}$ \eqref{eq:Q} 
and the time delays $\Delta t_\alpha$ \eqref{eqn:delay} vary with
angle, but the effect of $\Delta t_\alpha$ on the cross-correlation
terms occurs on a smaller angular scale for most of the sky.  A
simple criteria then is to cover the sky with a maximum angular
mismatch defined by the longest detector baseline and the maximum
frequency.  As with all template placement problems we have some
freedom in how to produce this set, and chose a somewhat sub-optimal
but simple set of directions - a grid of approximately $10^4$ points
uniformly distributed in $\theta$ and $\phi\sin\theta$.

\subsection{Waveform normalization}

The DFM waveforms are linearly polarized, which means that the two 
polarizations are linearly dependent and so can be written in the form
\bea
h_+(t) 
  & = &  \cos(2\psi_s) \, h(t) \\
h_\times(t) 
  & = &  \sin(2\psi_s) \, h(t) \, ,
\eea
\noindent
where $\psi_s$ is the polarization angle.  As a result, the strain
signal $g_\alpha$ in detector $\alpha$ is
\be\n{eqn:strainsignal}
g_\alpha(t)
  =  \left(
         F^+_\alpha(\widehat{\Omega}_s)\cos 2\psi_s 
        + F^\times_\alpha(\widehat{\Omega}_s)\sin 2\psi_s
     \right) h(t) \, .
\ee
\noindent
These waveforms are de-trended and normalized against the
interferometer noise curve so that $\rho^2_{\alpha \alpha}=1$
\eqref{eqn:SNR} for optimal orientation (the case 
$F^+_\alpha\cos 2\psi_s + F^\times_\alpha\sin 2\psi_s = 1$, so that
$g_\alpha(t)=h(t)$); i.e., we define the normalization of $h(t)$ so that
\be\n{eq:norm}
\sum_{k=0}^{N-1} \frac{|{\tilde h}[k]|^2}{{ \frac{N}{2} } S[k]} = 1 \, .
\ee
(Recall that our detectors have identical noise spectra, 
so $S_\alpha[k] = S_\beta[k] \equiv S[k]$.)  
The cross-correlations
\eqref{eqn:crossSNR} of the waveforms depend on their relative time or
phase shift; with this normalization the maximum cross-correlations
over all shifts for co-aligned detectors are
\be {\rm max}_{\Psi} \left\{ \rho_{\alpha\beta}^2 \right\} = \left\{
  \begin{array}{l l}
    0.58  &  \mathrm{(A1B3G3-A1B3G5)} \\
    0.26 & \mathrm{(A1B3G3-A3B4G2)} \\ 
    0.50 & \mathrm{(A1B3G5-A3B4G2)}
     \end{array} \right. ~ .
\ee
\noindent
For comparison, typical cross-correlation values for 
Gaussian noise in our time-frequency band are $0.15-0.2$.

To simulate a gravitational wave signal, one of the three waveforms
A1B3G3, A1B3G5, or A3B4G2 is randomly selected.  Uniformly distributed
random direction $\widehat{\Omega}_s$ and polarization
$\psi_s$ angles are chosen.  For each detector $\alpha$, the discrete
catalog waveform $h(t)$ is time-shifted by $\Delta t_\alpha
(\hat{\Omega}_s)$ \eqref{eqn:delay} and resampled to match
$n_\alpha[j]$.  The waveform is then scaled by the
antenna response as in \eqref{eqn:strainsignal} to give
$g_\alpha$.

To characterize the efficacy of our consistency test as a function of
the signal strength, we choose to simulate populations of candidates
with the same \emph{measured} signal-to-noise ratios; i.e., the
signals are scaled so as to deliver a fixed total SNR to the
network.  This simulates candidates near a detection threshold, rather
than (for example) a physical population of standard candles.  
With the normalization \eqref{eq:norm} and \eqref{eqn:strainsignal} 
one finds
\bea
\lefteqn{\frac{1}{D} \sum_{\alpha=1}^D \rho^2_{\alpha \alpha}
    =    \frac{1}{D} \sum_{\alpha=1}^D \sum_{k=0}^{N-1} 
         \frac{|g_{\alpha}[k]|^2}{{ \frac{N}{2} } S[k]}
}
         \nonumber \\
  & = &  \frac{1}{D} \displaystyle{\sum_{\alpha=1}^D} \left(
                 F^+_\alpha(\widehat{\Omega}_s)\cos2\psi_s
                 + F^\times_\alpha(\widehat{\Omega}_s)\sin2\psi_s
             \right)^2 \, .
\eea
To fix the SNR in the detectors to some
rms value $\rho_{\textrm{rms}}$, we apply the further normalization
\begin{equation}
g_\alpha(t) 
  \rightarrow 
    \frac{g_\alpha(t) \, \rho_{\textrm{rms}}}{
      \sqrt{\frac{1}{D} \displaystyle{\sum_{\beta=1}^D} \left( 
        F^+_\beta(\widehat{\Omega}_s)\cos 2\psi_s 
        + F^\times_\beta(\widehat{\Omega}_s)\sin 2\psi_s
      \right)^2}
    } \, .
\end{equation}
With this scaling we find
\be
\sqrt{\frac{1}{D} \sum_{\alpha=1}^D \rho^2_{\alpha \alpha}}
  =  \rho_{\textrm{rms}} \, .
\ee

To simulate a population of glitches, the same process was followed with
the sole exception that a different DFM waveform
was selected for each detector.  We
applied the same scaling and time delays as for a GWB.  This
population of glitches has the property that it would pass any
incoherent test, as its arrival times and power distribution are 
consistent with those of true gravitational-wave
signals.  We reiterate that, though not a realistic glitch population,
this provides us with examples of the kind of pathological glitches
that cannot be dismissed by per-detector methods.

\subsection{Analysis Procedure}

After the signal has been added to the background noise, the data are
whitened to produce $d_{\textrm{w}\alpha}[j]$.  The whitening algorithm 
is trained on a 16 second block of data that does not include the signal.
With a known set of trial sky positions $\{\widehat{\Omega}\}$ and
measured power spectra $S_\alpha(f)$, $\mathbf{Q}$ can be
computed for each direction and resolvable frequency.

For each direction on the sky, overlapping segments of data are
considered sequentially, the length of the segments depending on the
time scales of the signal under consideration (here chosen as $1/16$
s).  To perform the time shifts, the segments are extracted from the
closest integer samples in the time domain, and then transformed to
the Fourier domain where the remaining part of the time shift is
performed by applying phase shifts.  The null and incoherent energies
are then computed and recorded for that direction on the sky. 

For each simulation we select the two directions for which the
transient shows the most correlation according to two different
criteria, as well as the direction with the minimum null energy:
\begin{enumerate}
\item ${\rm min} (E_{\textrm {null}} - E_{\textrm {inc}})$: Measures
  the linear distance away from the diagonal in a scatter plot of
  $E_{\textrm {inc}}$ vs. $E_{\textrm {null}}$ (see Figure ~\ref{fig:inc_vs_null}).
  Physically it represents the largest amount of energy canceled in
  forming the null stream.
\item ${\rm min} (E_{\textrm {null}}/E_{\textrm {inc}})$: Measures the
  angular distance away from the diagonal in a scatter plot of
  $E_{\textrm {inc}}$ vs. $E_{\textrm {null}}$ (see Figure ~\ref{fig:inc_vs_null}).
  Physically it represents the largest fraction of energy canceled in
  forming the null stream.
\item ${\rm min} (E_{\textrm {null}})$: Calculated for comparison against
  the other two statistics, and for estimating the sky location.
\end{enumerate}

\begin{figure}
\resizebox{\columnwidth}{!}{\includegraphics{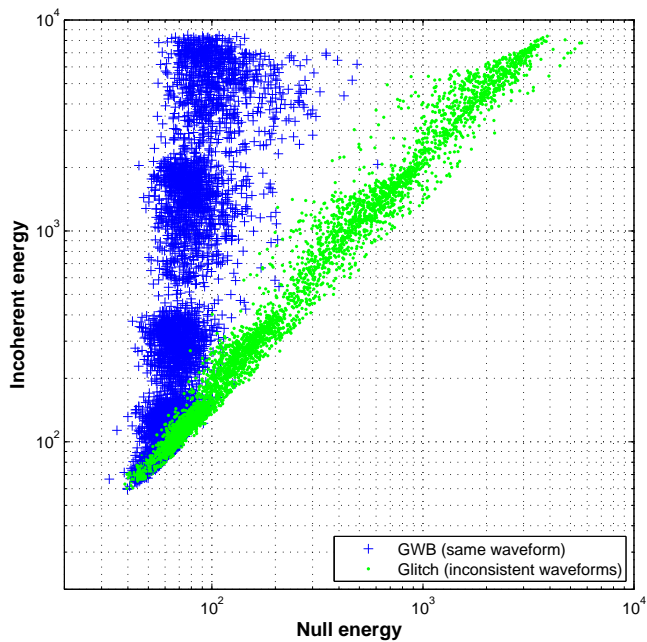}}
\caption{
  Scatter plot of the null and incoherent energies for the most
  correlated direction on the sky, defined as the direction of
  ${\textrm {min}} (E_{\textrm{null}}/E_{\textrm{inc}}$) (i.e., the upper
  left limit of Fig.~\ref{fig:inc_vs_null}), for glitches and GWBs of
  various signal-to-noise ratios.  Note that as the signal-to-noise
  ratio increases (higher $E_{\textrm{inc}}$) the GWB and glitch
  distributions separate, with the GWBs remaining at relatively low
  null energies and the glitches having comparable null and incoherent
  energies.  The clumping is due to the 5
  distinct rms SNRs used in our simulations: $5, 10, 20, 50, 100$.
  The null energies for some of the GWB simulations are greater than that 
  expected (60) from the number of degrees of freedom because we are selecting sky
  positions from the minimum of $E_{\textrm{null}}/E_{\textrm{inc}}$, not
  from the minimum of $E_{\textrm{null}}$.
}
\label{fig:scatter_null_over_inc}
\end{figure}

\begin{figure}
\resizebox{\columnwidth}{!}{\includegraphics{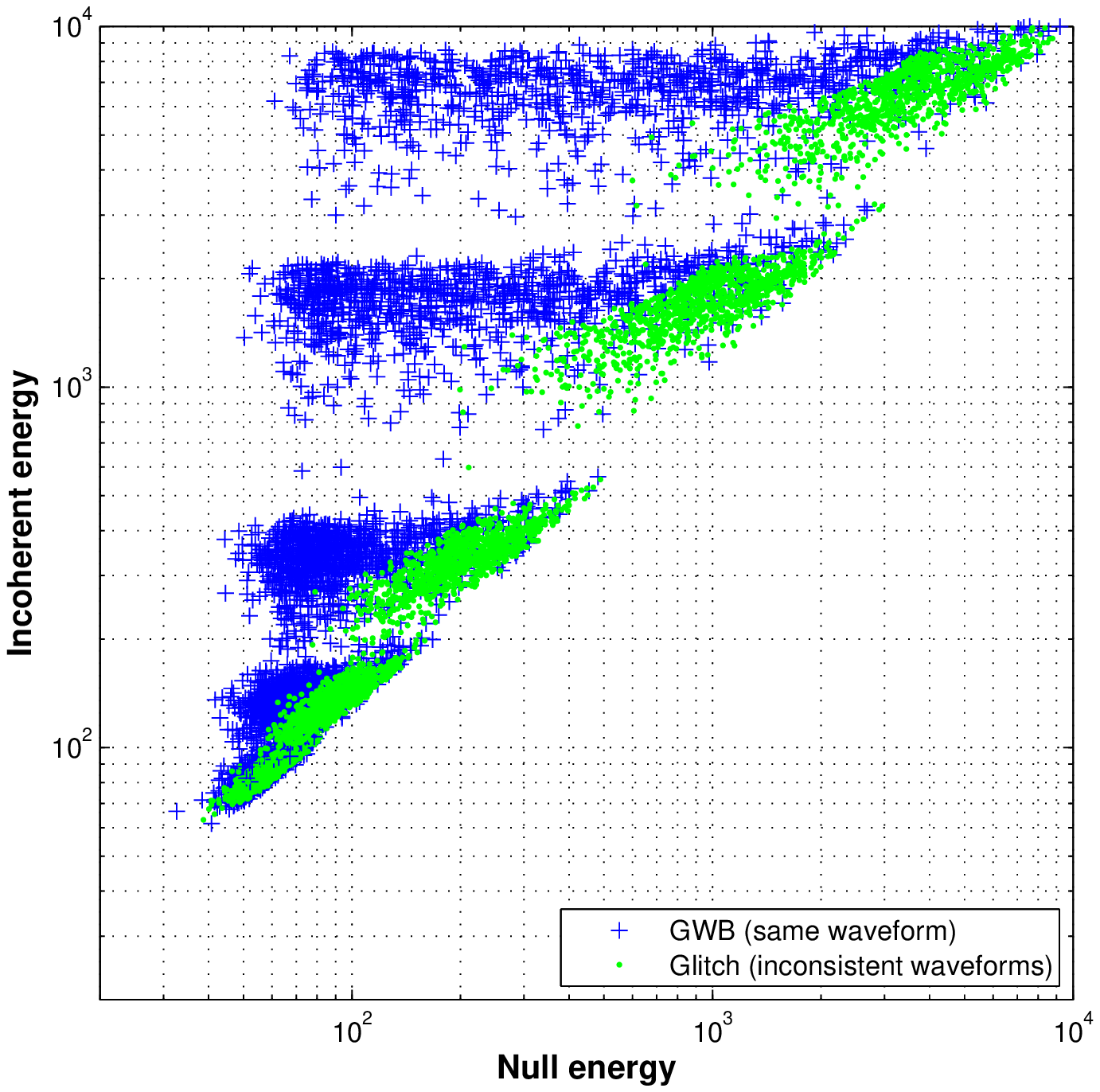}}
\caption{
  Scatter plot of the null and incoherent energies for the most
  correlated direction on the sky, defined as the direction of
  max($E_{\textrm{inc}}-E_{\textrm{null}}$) (i.e., the upper left limit of
  Fig.~\ref{fig:inc_vs_null}), for glitches and GWBs of various
  signal-to-noise ratios.  Note that as the signal-to-noise ratio
  increases (higher $E_{\textrm{inc}}$) the GWB and glitch distributions
  separate, with the GWBs remaining at relatively low null energies
  and the glitches having comparable null and incoherent energies.
  The clumping into horizontal bands is due to the 5 distinct rms SNRs
  used in our simulations: $5, 10, 20, 50, 100$.  The null energies
  for many of the GWB simulations are greater than that expected (60) from the number of
  degrees of freedom because we are selecting sky positions from the
  minimum of $E_{\textrm{null}}-E_{\textrm{inc}}$, not from the minimum of
  $E_{\textrm{null}}$. 
}
\label{fig:scatter_null-inc}
\end{figure}

\begin{figure}
\resizebox{\columnwidth}{!}{\includegraphics{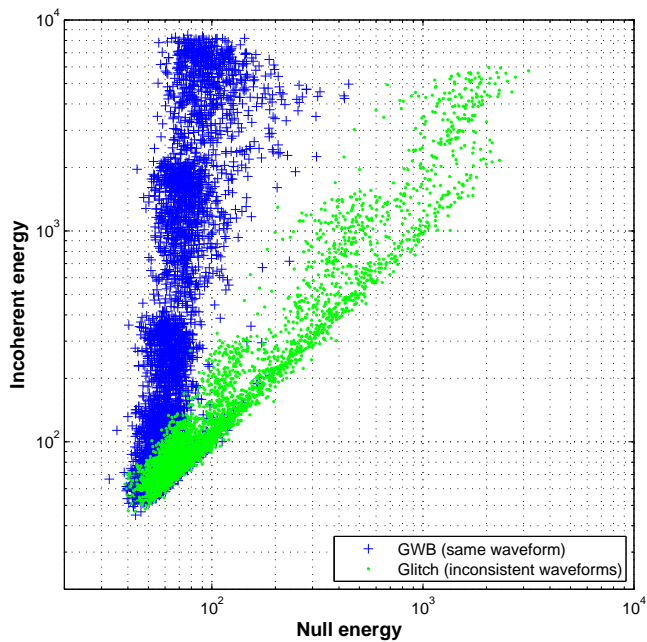}}
\caption{
  Scatter plot of the null and incoherent energies for the best-fit
  direction on the sky, defined as the direction of 
  ${\rm {min}}(E_{\textrm{null}})$ (i.e., the left-most point in 
  Figure~\ref{fig:inc_vs_null}),
  for glitches and GWBs of various signal-to-noise
  ratios.  Note that as the signal-to-noise ratio increases (higher
  $E_{\textrm{inc}}$) the GWB and glitch distributions separate, with the
  GWBs remaining at relatively low null energies and the glitches
  having comparable null and incoherent energies.  The clumping
  is due to the 5 distinct rms SNRs used in our
  simulations: $5, 10, 20, 50, 100$.  Note that a large fraction of
  the glitch signals produce low null energies for some sky positions.
  This means that GWBs and glitches are not distinguishable using only
  the null energy. 
}
\label{fig:scatter_null}
\end{figure}

\begin{figure}
\resizebox{\columnwidth}{!}{\includegraphics{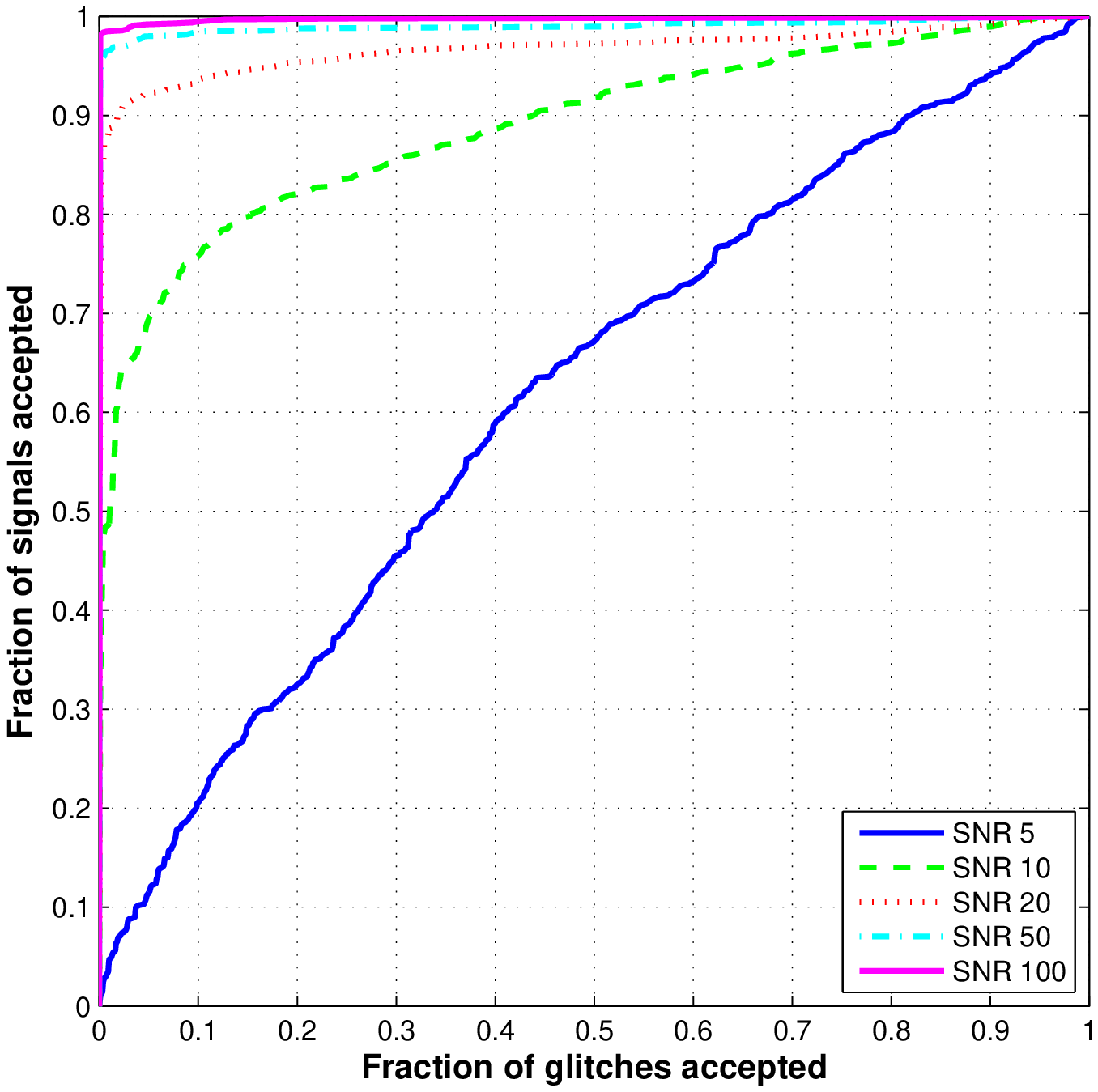}}
\caption{
  Receiver-Operator Characteristic (ROC) plot for ${\rm min} (E_{\textrm
    {null}}/E_{\textrm {inc}})$ as a statistic for distinguishing GWBs
  from noise glitches.  The ROC curve value is given by the fraction of GWBs
  (true acceptance) and glitches (false acceptance) of given rms SNR
  falling to the left of a line of constant $E_{\textrm{null}}/E_{\textrm{inc}}$ 
  (a diagonal line) in Fig.~\ref{fig:scatter_null_over_inc}.
  The rapid rise of the curves at low false
  acceptance is indicative of the ability of the method to confidently
  distinguish a significant portion of GWB signals from the glitch
  population.  
}
\label{fig:ROC_null_over_inc}
\end{figure}

\begin{figure}
\resizebox{\columnwidth}{!}{\includegraphics{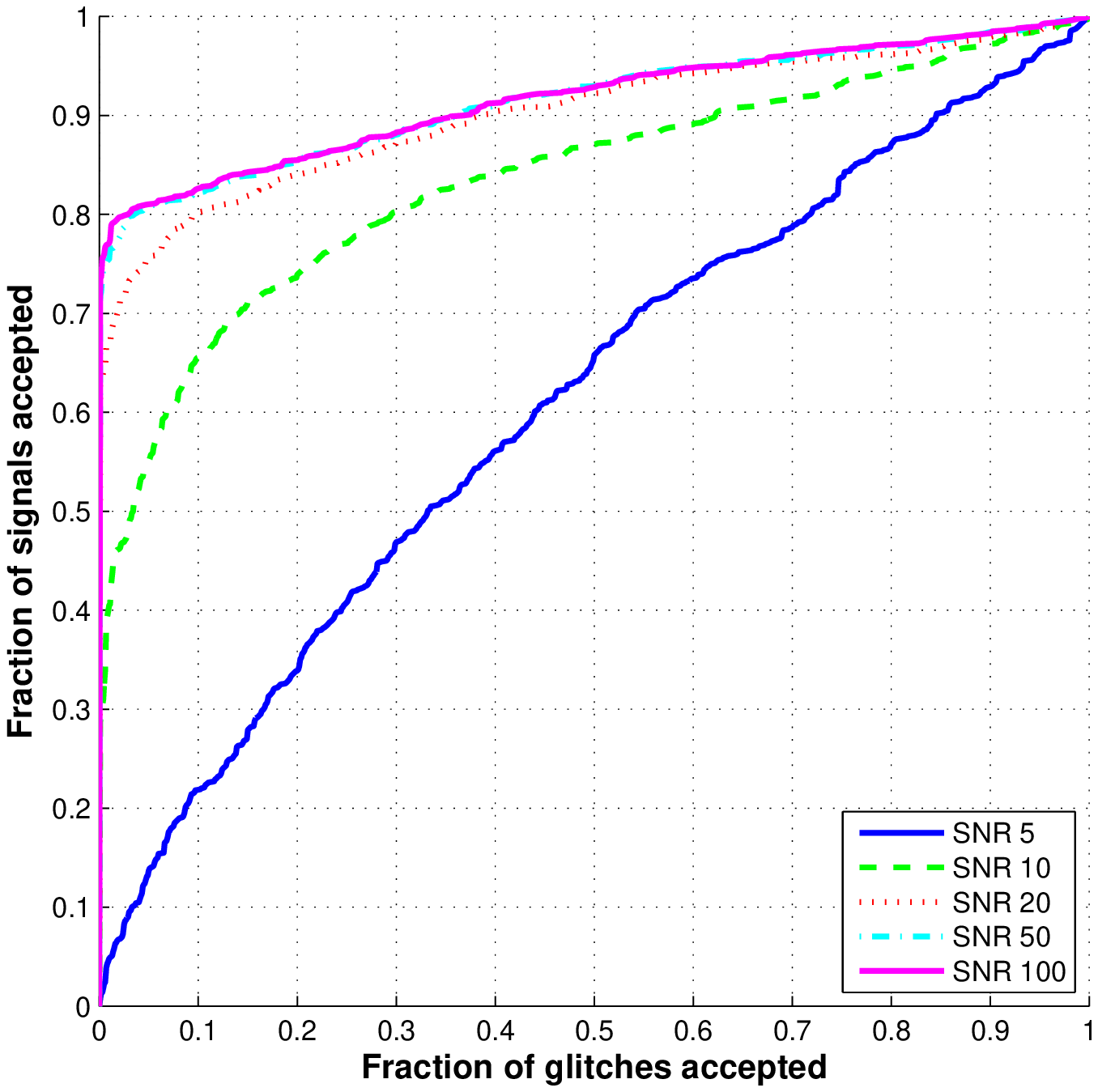}}
\caption{
  Receiver-Operator Characteristic plot (ROC) for ${\rm {min}}
  (E_{\textrm{null}} - E_{\textrm{inc}})$ as a statistic for distinguishing
  GWBs from noise glitches.  The ROC curve value is given by the
  fraction of GWBs (true acceptance) and glitches (false acceptance)
  of given rms SNR falling to the left of a line of
  constant $E_{\rm null}-E_{\rm inc}$ in Fig.~\ref{fig:scatter_null-inc}.
  The rapid rise of the curves at low false
  acceptance is indicative of the ability of the method to confidently
  distinguish a significant portion of GWB signals from the glitch
  population. This is not as powerful a statistic as ${\rm {min}}
  (E_{\textrm{null}}/E_{\textrm{inc}})$ (see Figure
  ~\ref{fig:ROC_null_over_inc}).  
}
\label{fig:ROC_null-inc}
\end{figure}

\begin{figure}
\resizebox{\columnwidth}{!}{\includegraphics{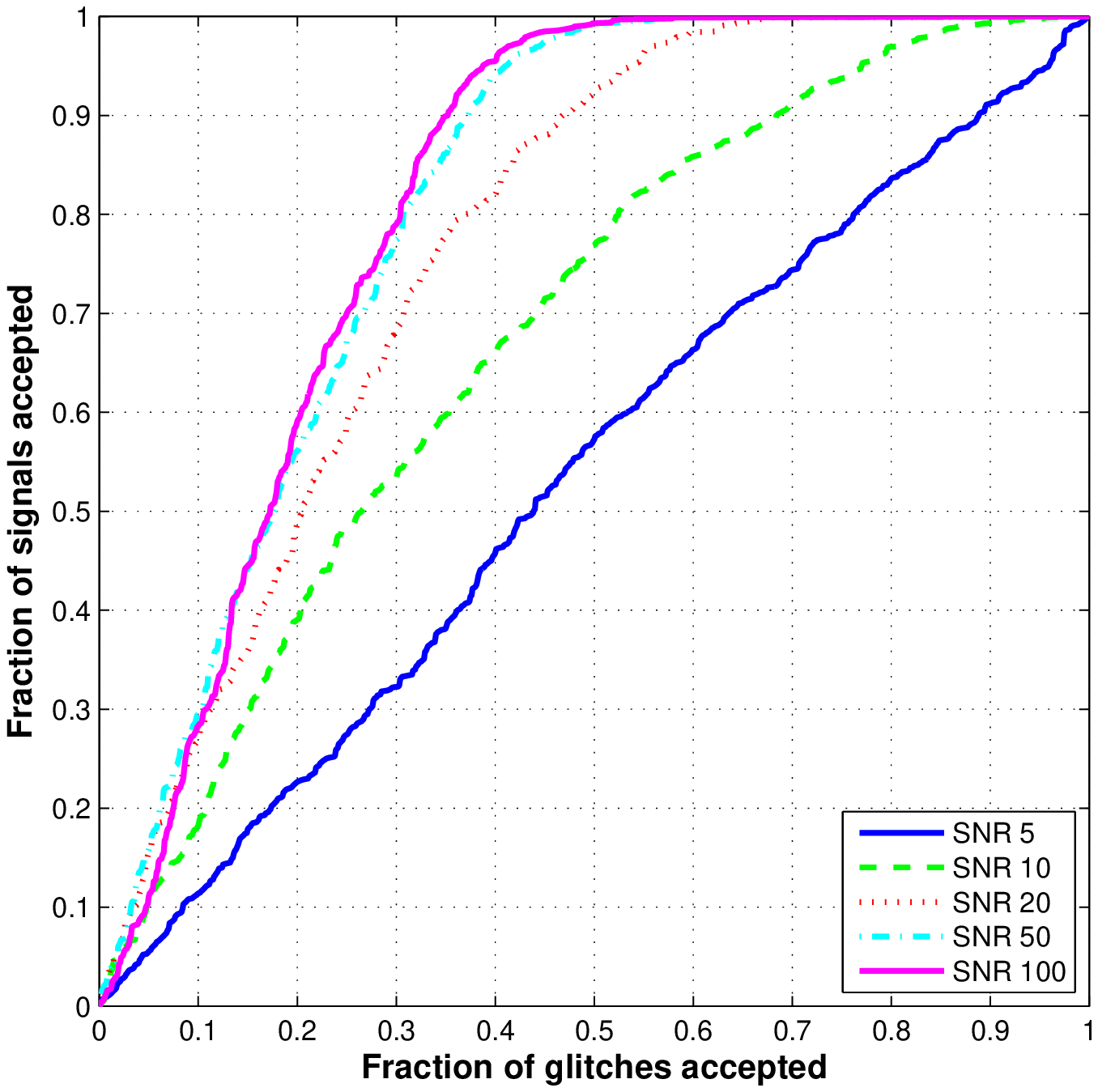}}
\caption{
  Receiver-Operator Characteristic (ROC) plot for ${\rm min}(E_{\rm null})$
  as a statistic for distinguishing GWBs from noise glitches. 
  The ROC curve value is given by the fraction of
  GWBs (true acceptance) and glitches (false acceptance) of given rms
  SNR falling to the left of a line of constant 
  $E_{\rm null}$ (a vertical line) in Fig.~\ref{fig:scatter_null}.  
  The lower slope of the curves at low acceptance is due to the fact that
  many glitches have sky positions for which the null energy is small,
  even when the SNR of the glitch is high. The null energy alone is
  therefore not very effective for confidently distinguishing GWBs
  from noise glitches.  While one could chose a threshold that varies with 
  $E_{\rm inc}$ to get better performance, ${\rm {min}}(E_{\textrm{null}})$
  is still not as powerful a statistic as 
  ${\rm {min}}(E_{\textrm{null}}/E_{\textrm{inc}})$.
}
\label{fig:ROC_null}
\end{figure}

Figures~\ref{fig:scatter_null_over_inc},~\ref{fig:scatter_null-inc},~\ref{fig:scatter_null} show scatter plots
of $E_{\textrm {inc}}$ vs. $E_{\textrm {null}}$ for the sky locations
picked by these three criteria for $10^4$ simulated signals. The
signal population consists of $10^3$ GWBs and $10^3$ glitches at each
of $5$ different signal-to-noise ratios, $\rho_{\rm {rms}} = 5, 10,
20, 50, 100$.

Note that there is a significant difference in the distributions of
signals and glitches using any of the measures 
${\rm min}(E_{\textrm{null}}-E_{\textrm {inc}})$, 
${\rm min}(E_{\textrm{null}}/E_{\textrm {inc}})$, or 
${\rm min}(E_{\textrm {null}})$.
For both GWBs and glitches there are directions on the sky for which
the null energy is low.  However, for glitches these occur exclusively
at low incoherent energies.  For signals there exist directions with
low null energy at larger incoherent energies. This restriction is
strongest when measuring correlations using 
${\rm min}(E_{\textrm{null}}/E_{\textrm {inc}})$ 
(Figure~\ref{fig:scatter_null_over_inc}). For 
${\rm min}(E_{\textrm {null}}-E_{\textrm {inc}})$ 
(Figure~\ref{fig:scatter_null-inc}) there are a
small but noticeable fraction of GWBs that overlap the glitch
population and so are indistinguishable from glitches. By using
instead ${\rm min}(E_{\textrm {null}})$ only 
(Figure~\ref{fig:scatter_null}) a significant fraction of the glitches 
have a sky location with low null energy, but only at low incoherent 
energy as well.

As the SNR increases the populations become distinct.  We can see this
in Figures~\ref{fig:ROC_null_over_inc}-\ref{fig:ROC_null}, which show the 
Receiver-Operator Characteristic (ROC) curves for the performance 
of our various statistics in discriminating the two populations.  
At total energies corresponding to
$\rho_{\rm rms} \sim 10 - 20$ or greater we can detect most of a
population of gravitational waves and reject essentially all of a population of
semi-correlated glitches.  The rejected gravitational waves are those
that are weak in at least one detector, or equivalently produce little
correlation.

For $\rho_{\rm rms} = 5$ or lower the glitch and GWB populations are
not distinct. This is because the null stream enforces waveform
consistency only when there is excess energy to suppress.

\section{Conclusions and Future directions}
\label{SECIV}

In this article we have introduced an extension of the
G\"ursel-Tinto null-stream technique that allows one to make robust
tests to distinguish between GWB signals and coincident noise glitches.  
This technique is based on comparing the energy in the null stream 
to that expected if the signals in the various detectors are 
uncorrelated, and does not require any {\em a priori} knowledge 
of the GWB or glitch waveforms.  We applied this technique to the 
case of the LIGO-Virgo 3-detector network at design sensitivity, 
and quantified the ability of three different measures of 
correlation to distinguish true GWBs from coincident noise glitches. 
For the best-performing measure, the ratio of null energy to 
incoherent energy, we found that gravitational-wave bursts of 
SNR $10-20$ or greater can be distinguished from
glitches of comparable SNRs that are injected in the data with the
same time delays but are different in the three detectors' data.
For example, with the GWB and glitch populations tested, we 
found that 90\% of glitches can be rejected while accepting 94\% of 
SNR 20 GWBs and 76\% of SNR 10 GWBs.  Furthermore, we stress that the 
glitch population tested was pathological in the sense that 
they were constructed to have time delays and amplitudes consistent 
with a GWB.  Hence, the performance of the consistency test may be 
even better with real detector data.  This consistency test is 
therefore a promising technique for rejecting noise coincidences 
and increasing detection confidence in GWB searches.

The development of coherent analysis techniques for GWB detection is
still at an early stage, and much further research can be done.  In this
section we briefly outline some of the directions of current and
future work.  These can be divided roughly into applications of the
existing consistency test to more general networks and signals, and
extensions and improvements to the algorithm.

We will systematically study a larger variety of
waveforms than the small set considered in the paper.  These should
include two-polarization GWBs.  (The supernova waveforms used here are
linearly polarized).  This wider set may include various supernova
catalogs \cite{ZwMu:97,DiFoMu:02b,OtBuLiWa:04}, and approximate
waveforms for black-hole binary coalescence (see for example 
\cite{Pr:05}).  The latter are particularly important, since it 
is quite plausible that black-hole binaries will be the first 
transient signals to be detected.

Another near-term goal is to apply our modified null-stream test to
other networks, such as those with four and five detectors.  A fourth
non-aligned detector will reduce the fraction of sky over which only
two detectors have significant sensitivity \cite{Tinto96}, increasing
the strength of the test.  An additional aligned detector, such as the
two-kilometer detector at LIGO-Hanford, would also provide a second
null stream without extra sky coverage.  Lazzarini {\em et al.\/}
\cite{AL:04} have demonstrated how the output of the two LIGO-Hanford
detectors can be combined to form a single pseudo-detector with
greater sensitivity than either.  The difference between the detector
outputs is also a null stream, effectively a detector with zero
antenna response, which can be employed in our consistency test.  This
is a computationally cheap test since the H1-H2 null stream is
independent of the sky position, providing the basis of a simple
hierarchical analysis scheme.

We also plan to test the power of our consistency test on real data, with
real noise transients.  We should note that the artificial noise
coincidences studied in this work are pathological in the sense that
they are injected with time delays and amplitude responses consistent
with actual sky positions, and waveforms that are identical (in
individual detectors) to GWBs.  Although we would not expect
noise in actual detectors to be so pathological, it is important to
characterize the robustness of our technique for real data, and this
was the motivation for our implementation.

Another important aspect of real data is that their calibration could
be inaccurate, implying an imperfect cancellation of the GWB by the
null stream.  We will study quantitatively the effects of realistic
calibration errors on the effectiveness of the null-stream technique
(see also the study by Ajith {\em et al.\/} \cite{AjHeWeHe:05}).

A first improvement is to optimize the algorithm for implementing our test.
An optimal identification of the integration time and frequency band over
which the null and incoherent energies are calculated will improve the 
effectiveness of the test by minimizing the amount of noise included. 

An important augmentation of the coherent analysis is improved
techniques for determining the sky location of the source.  
Current efforts estimate the sky position as the extremum in a sky map 
of the null stream or some likelihood statistic \cite{GuTi:89,Klimenko:05}. 
For example, G\"usel and Tinto \cite{GuTi:89} demonstrated that 
locating the minimum of the null energy allows one to determine 
the direction to the source of a kiloHertz GWB with high accuracy 
for typical SNRs of $\sim 40-60$ (converted to our $\rho_{\textrm{rms}}$). 
However, as we have seen, sky maps exhibit
structure which is a combination of the network geometry and the
signal waveform.  For example, linearly polarized signals produce
interference fringes in the sky map.  These fringes appear as rings 
of fixed time delay with
respect to the various detector baselines. A global analysis which
takes account of this structure could in principle average over local
noise fluctuations in the sky map to achieve an improved pointing
accuracy for weaker signals.

We will further investigate Bayesian interpretations and formulations
of the null-stream technique, and compare the effectiveness of this
approach against the procedure presented in this work. We will also
investigate the possibility of incorporating 
additional tests such as distribution-free
(non-parametric) correlation tests. These could prove valuable
when analyzing real data since they enforce known statistics even when
the noise in the data follows an arbitrary noise distribution function.
 
\appendix

\section{Null Stream Projection Operator}
\label{sec:projection}

In this section we derive explicit expressions for 
the projection operator which acts on the
network data vector to produce null streams.  This projection operator
projects the data orthogonally to $\boldsymbol{F_{\textrm{w}}^+}$ and 
$\boldsymbol{F_{\textrm{w}}^\times}$.

Let $\boldsymbol{m}$ and $\boldsymbol{n}$ be any orthonormal pair of 
vectors that span the space spanned by $\boldsymbol{F_{\textrm{w}}^+}$ and 
$\boldsymbol{F_{\textrm{w}}^\times}$ (the vectors 
$\boldsymbol{F_{\textrm{w}}^+}$ and $\boldsymbol{F_{\textrm{w}}^\times}$ 
are not necessarily orthogonal).  Then the projection operator that removes
the gravitational-wave contribution is
\be\label{eqn:PNS}
P^{\textrm{NS}}_{\alpha\gamma}
  =  \sum_{\beta = 1}^D
     (\delta_{\alpha\beta} - m_\alpha m_\beta)
     (\delta_{\beta\gamma} - n_\beta n_\gamma) \, .
\ee
For example, choosing
\bea
\boldsymbol{m}
  & = &  \frac{\boldsymbol{F_{\textrm{w}}^+}}{|\boldsymbol{F_{\textrm{w}}^+}|}
          \, , \label{eqn:mvector} \\
\boldsymbol{n}
  & = &  \frac{
             \boldsymbol{F_{\textrm{w}}^\times}
             -(\boldsymbol{m}\cdot\boldsymbol{F_{\textrm{w}}^\times})\boldsymbol{m}
         }{
             |\boldsymbol{F_{\textrm{w}}^\times}
             -(\boldsymbol{m}\cdot\boldsymbol{F_{\textrm{w}}^\times})\boldsymbol{m}|
         } \, , \label{eqn:nvector} 
\eea
one finds
\be\label{eqn:PNSsoln}
P^{\textrm{NS}}_{\alpha\gamma}
  =  \delta_{\alpha\gamma}
     - \frac{1}{\det(M)}\left[\begin{array}{c c}
         F_{\textrm{w}\alpha}^+  \!& \! F_{\textrm{w}\alpha}^\times
     \end{array}\right] \!\!
     \left[\begin{array}{r r}
         M_{\times\times} \! & \! -M_{+\times}\\
         -M_{\times+}     \! & \!  M_{++}\\
     \end{array}\right] \!\!
     \left[\begin{array}{c}
         F_{\textrm{w}\gamma}^+ \\ F_{\textrm{w}\gamma}^\times
     \end{array}\right]
\ee
where the $2\times2$ matrix $M$ is defined by
\be\label{eqn:Mmatrix}
\boldsymbol{M} 
  = \boldsymbol{F_\textrm{w}}^T \boldsymbol{F_\textrm{w}} \, .
\ee
In matrix notation \eqref{eqn:PNSsoln} becomes 
\be
\boldsymbol{P}^{\textrm{NS}} 
  =  \boldsymbol{I} 
     - \boldsymbol{F_\textrm{w}} \boldsymbol{M}^{-1} \boldsymbol{F_\textrm{w}}^T 
     \, ,
\ee

If $\boldsymbol{F_{\textrm{w}}^+} \propto \boldsymbol{F_{\textrm{w}}^\times}$ 
(e.g., for co-aligned detectors) then the projection operator simplifies to
\be\label{eqn:PNSaligned}
P^{\textrm{NS}}_{\alpha\beta}
  =  (\delta_{\alpha\beta} - m_\alpha m_\beta) 
\ee
with $m_\alpha$ given by \eqref{eqn:mvector}.

\section{Minimum-Variance Waveform Reconstruction}
\label{sec:reconstruction}

In this section we derive explicit expressions for the two amplitude
components of the wave, $h_+$ and $h_\times$, as optimally
reconstructed as linear combinations of the detector data.
This is a generalization of the technique used by G\"ursel and 
Tinto \cite{GuTi:89}, and was first derived by 
Flanagan and Hughes \cite{FlHu:98b}.

We will assume that the whitened data streams have been time-shifted
before these combinations are constructed, and will simply write
$\tilde{d}_{\textrm{w}\alpha}[f_i]$ for 
\mbox{$\tilde{d}_{\textrm{w}\alpha}[f_i]\e^{i2\pi f_i\Delta t_\alpha(\theta,\phi)}$}, 
$\alpha = 1, \ldots D$.

Our goal can be reduced to the problem of identifying two vectors
$\boldsymbol{V}_+$, $\boldsymbol{V}_\times$ such that
\bea
\tilde{h}_+^{\textrm{est}}
  & \equiv &  \boldsymbol{V}_+ \cdot \boldsymbol{\tilde{d}_{\textrm{w}}}  =  
\tilde{h}_+ + \boldsymbol{V}_+ \cdot \boldsymbol{\tilde{n}_{\textrm{w}}}  \\
\tilde{h}_\times^{\textrm{est}}
  & \equiv &  \boldsymbol{V}_\times \cdot \boldsymbol{\tilde{d}_{\textrm{w}}}  =    
\tilde{h}_\times + \boldsymbol{V}_\times \cdot \boldsymbol{\tilde{n}_{\textrm{w}}} \ .
\eea
From the above equations we conclude that these two vectors must satisfy 
the constraints
\bea
\boldsymbol{V}_+ \cdot \boldsymbol{F_{\textrm{w}}^+}      & = & 1
\nonumber
\\
\boldsymbol{V}_+ \cdot \boldsymbol{F_{\textrm{w}}^\times} & = & 0 
\nonumber
\\
\boldsymbol{V}_\times \cdot \boldsymbol{F_{\textrm{w}}^+}      & = & 0 
\nonumber
\\
\boldsymbol{V}_\times \cdot \boldsymbol{F_{\textrm{w}}^\times} & = & 1 \ .
\label{eqn:Vconstraints}
\eea
This implies that the errors in estimating the
two waveforms are
\bea
\delta \tilde{h}_+^{\textrm{est}}
  & = &  \tilde{h}_+^{\textrm{est}} - \tilde{h}_+  
    =    \boldsymbol{V}_+ \cdot \boldsymbol{\tilde{n}_{\textrm{w}}} \\
\delta \tilde{h}_\times^{\textrm{est}}
  & = &  \tilde{h}_\times^{\textrm{est}} - \tilde{h}_\times
    =    \boldsymbol{V}_\times \cdot \boldsymbol{\tilde{n}_{\textrm{w}}} 
         \, .
\label{Errors_h}
\eea

If the noises affecting the detectors are Gaussian distributed, the
``optimal'' choice for $\boldsymbol{V}_+, \boldsymbol{V}_\times$ is
that implied by minimizing the mean-square errors of the noises
affecting the two reconstructed waveforms,
\bea
\frac{1}{N} \sum_{k=0}^{N-1} |\delta \tilde{h}_+^{\textrm{est}}[k]|^2
  & = &  \sum_{\alpha=1}^D V^{+2}_\alpha  \, , \\
\frac{1}{N} \sum_{k=0}^{N-1} |\delta \tilde{h}_\times^{\textrm{est}}[k]|^2
  & = &  \sum_{\alpha=1}^D V^{\times2}_\alpha \, . 
\eea
That is, we want the shortest (in the Cartesian sense) vectors 
$\boldsymbol{V}_+, \boldsymbol{V}_\times$ which satisfy the 
constraints (\ref{eqn:Vconstraints}).
This minimization can be performed by using the method of the Lagrange
multipliers. For instance, for $\boldsymbol{V}_+$ we have
\bea 
0 & = &  \frac{\delta}{\delta V^+_\alpha} \left[ \, 
             \sum_\beta V^{+2}_\beta 
             + \lambda_+ \left(
                 \sum_\beta V^+_\beta F^+_{\textrm{w}\beta} - 1 
             \right)
             \right. \nonumber \\
  &   &      \left. \mbox{} 
             + \lambda_\times \left( 
                 \sum_\beta V^+_\beta F^\times_{\textrm{w}\beta} 
             \right) 
         \, \right] \, , 
\eea
where $(\lambda_+, \lambda_\times)$ are the two Lagrange multipliers.
The above equation leads to 
\be
V^+_\alpha
  =  - \frac{1}{2}\left(
         \lambda_+ F_{\textrm{w}\alpha}^+ 
         + \lambda_\times F_{\textrm{w}\alpha}^\times
     \right)  \, .
\ee
Imposing the constraints (\ref{eqn:Vconstraints}) in the above
equation results in the linear system of equations
\be
\left(\begin{array}{r}
    -2 \\
     0
\end{array}\right)
  =  M \left(\begin{array}{r}
         \lambda_+ \\
         \lambda_\times
     \end{array}\right)
     \, ,
\ee
where the symmetric $2\times2$ matrix $M$ is defined in \eqref{eqn:Mmatrix}. 
Inverting $\boldsymbol{M}$, it follows that
\be
\left(\begin{array}{c}
    \lambda_+ \\
    \lambda_\times
\end{array}\right)
  =  \frac{2}{\det(M)}\left(\begin{array}{r}
         -M_{\times\times} \\
         M_{+\times}
     \end{array}\right) \, ,
\label{lambdaPC}
\ee
\noindent
which gives 
\be\n{eqn:Vpw}
V^+_\alpha
   =  \frac{1}{\det(M)}\left(
          F_{\textrm{w}\alpha}^+ M_{\times\times} \\
          - F_{\textrm{w}\alpha}^\times M_{+\times}
      \right) \, .
\ee
By performing an analogous calculation for $V^\times_\alpha$ we find
\be\n{eqn:Vcw}
V^\times_\alpha
  =  \frac{1}{\det(M)}\left(
         - F_{\textrm{w}\alpha}^+ M_{+\times} \\
         + F_{\textrm{w}\alpha}^\times M_{++}
     \right)  \, .
\ee
\noindent
Note that the weighting of a particular detector $d_\alpha$ vanishes
if $F_{\textrm{w}\alpha}^+=0=F_{\textrm{w}\alpha}^\times$, which occurs if
detector $d_\alpha$ has no sensitivity to the sky location being considered
($F^+_\alpha=0=F^\times_\alpha$), or if it is much noisier
than the other detectors ($S_\alpha\to\infty$).  In these cases,
the expressions (\ref{eqn:Vpw},\ref{eqn:Vcw}) at that frequency and 
sky position reduce to those for the
network that does not include detector $\alpha$.

In matrix form, $V^{+,\times}_\alpha$ are the rows of a $2\times D$ matrix 
$\boldsymbol{V}$ where from \eqref{eqn:Vpw}, \eqref{eqn:Vcw}, 
and \eqref{eqn:Mmatrix}
\be
\boldsymbol{V}  =  \boldsymbol{M}^{-1} \boldsymbol{F}^T 
  =  (\boldsymbol{F}^T\boldsymbol{F})^{-1} \boldsymbol{F}^T \, .
\ee
That is, the waveform reconstruction matrix $\boldsymbol{V}$ is the 
{\em pseudo-inverse} of $\boldsymbol{F}$ \cite{Ra:06}.  
Indeed, it is easily verified that
\be\label{eqn:V}
\boldsymbol{V} \boldsymbol{F}  =  \boldsymbol{I} \, .
\ee
This is equivalent to the conditions \eqref{eqn:Vconstraints}.

If $\boldsymbol{F_{\textrm{w}}^+} \propto \boldsymbol{F_{\textrm{w}}^\times}$ 
(e.g., for co-aligned detectors) then only a single linear combination of the two polarizations 
$h_{+,\times}$ can be extracted.  Choosing the polarization gauge $\psi'$ such that 
$|\boldsymbol{F_{\textrm{w}}^{'+}}| = (|\boldsymbol{F_{\textrm{w}}^+}|^2+|\boldsymbol{F_{\textrm{w}}^\times}|^2)^{1/2}$, 
$\boldsymbol{F_{\textrm{w}}^{'\times}}=0$, the reconstruction matrix reduces to simply
\be\label{eqn:Valigned}
\boldsymbol{V}
  =  \frac{\boldsymbol{F_{\textrm{w}}^{'+}}^{T}}{|\boldsymbol{F_{\textrm{w}}^{'+}}|^2} \, .
\ee

See Rakhmanov \cite{Ra:06} for a discussion of singularities in waveform 
reconstruction due to rank-deficiency of $\boldsymbol{F_{\textrm{w}}}$, and the use of 
Tikhonov regularization to avoid such problems.  Rakhmanov's expressions 
correspond to ours with the replacement 
$F^{+,\times}_{\alpha}\to F^{+,\times}_{\textrm{w}\alpha}$; i.e., 
replacing the antenna responses by the noise-weighted antenna responses.

\begin{acknowledgments}
  
  This work was performed under partial funding from the following NSF
  Grants: PHY-0107417, 0140369, 0239735, 0244902, 0300609, and
  INT-0138459.  A.~Searle was supported by the Australian Research
  Council and the LIGO Visitors Program.  L.~Stein was supported in
  part by an NSF REU Site grant.  Simulations were performed on the
  Australian Partnership for Advanced Computing's National Facility
  under the Merit Allocation Scheme. For M.~Tinto, the research was
  also performed at the Jet Propulsion Laboratory, California
  Institute of Technology, under contract with the National
  Aeronautics and Space Administration.  This document has been
  assigned LIGO Laboratory document number LIGO-P060009-01-E.

\end{acknowledgments}

\appendix
\bibliography{master}
\end{document}